\documentclass[letterpaper, 11pt]{article}
\pdfoutput=1

\usepackage[nocompress]{cite}
\usepackage{jheppub}

\usepackage{graphicx}
\usepackage{epstopdf}
\usepackage{amsmath, amssymb}

\usepackage{bm}
\usepackage{bbold}

\newcommand{\be}{\begin{eqnarray}}
\newcommand{\ee}{\end{eqnarray}}
\newcommand{\nn}{\nonumber}
\newcommand{\bn}{\begin{enumerate}}
\newcommand{\en}{\end{enumerate}}

\def\IC{\mathbb{C}}

\def\IZ{\mathbb{Z}}

\def\CC{{\cal C}}

\def\CF{{\cal F}}

\def\CM{{\cal M}}
\def\CN{{\cal N}}

\def\CS{{\cal S}}

\def\CW{{\cal W}}

\def\a{\alpha}

\def\e{\epsilon}

\def\l{\lambda}

\def\s{\sigma}

\def\ch{\chi}
\def\G{\Gamma}

\def\half{\frac{1}{2}}

\def\Tr{{\rm Tr}}

\def\fp{\mathfrak{p}}
\def\fq{\mathfrak{q}}
\def\PE{\textrm{PE}}

\def\vec#1{\bm{#1}}

\title{Superconformal indices of generalized Argyres-Douglas theories from 2d TQFT}

\author{Jaewon Song}
\affiliation{Department of Physics, University of California, San Diego \\La Jolla, CA 92093, USA}
\emailAdd{jsong@physics.ucsd.edu}

\abstract
{
We study superconformal indices of 4d $\CN=2$ class $\CS$ theories with certain irregular punctures called type $I_{k, N}$. This class of theories include generalized Argyres-Douglas theories of type $(A_{k-1}, A_{N-1})$ and more. We conjecture the superconformal indices in certain simplified limits based on the TQFT structure of the class $\CS$ theories by writing an expression for the wave function corresponding to the puncture $I_{k, N}$. 
We write the Schur limit of the wave function when $k$ and $N$ are coprime. When $k=2$, we also conjecture a closed-form expression for the Hall-Littlewood index and the Macdonald index for odd $N$. From the index, we argue that certain short-multiplet which can appear in the OPE of the stress-energy tensor is absent in the $(A_1, A_{2n})$ theory. 
We also discuss the mixed Schur indices for the $\CN=1$ class $\CS$ theories with irregular punctures. 
}

\begin{document}
\maketitle

\section{Introduction}

Four-dimensional $\CN=2$ supersymmetric field theories in class $\CS$ refers to the ones that can be realized by wrapping 6d $\CN=(2, 0)$ theory on a Riemann surface $\CC$ \cite{Gaiotto:2009we, Gaiotto:2009hg}. This description enables us to understand dynamics of the 4d theory in terms of geometry of the Riemann surface. One of the most interesting connection is between the superconformal index \cite{Kinney:2005ej,Romelsberger:2005eg} of the 4d theory and the 2d topological field theory \cite{Gadde:2009kb, Gadde:2011ik,Gadde:2011uv, Gaiotto:2012xa,Rastelli:2014jja}. It says that the superconformal index of a given theory labelled by $\CC$ (called the UV curve) is given by a correlation function of the 2d topological field theory on $\CC$, which is a deformed version of Yang-Mills theory. Especially, in the Schur limit of the index, the TQFT is identified as the $q$-deformed Yang-Mills theory \cite{Aganagic:2004js}. It has been shown via localization of 5d Maximal SYM on $S^3$, that the 2d theory is indeed given by the $q$-deformed Yang-Mills theory \cite{Kawano:2012up, Fukuda:2012jr, Kawano:2015ssa}. This relation is also extended to the Lens space index \cite{Benini:2011nc, Alday:2013rs, Razamat:2013jxa}, to the outer-automorphsim twisted index \cite{Mekareeya:2012tn}, and to other gauge groups \cite{Lemos:2012ph, Chacaltana:2013oka, Agarwal:2013uga, Chacaltana:2014jba, Chacaltana:2015bna}. 

A class $\CS$ theory is not just labelled by the UV curve but also its local data on the punctures. There are regular and irregular punctures depending on the boundary condition we impose. The regular punctures are labelled by an SU(2) embedding into $\G$ that labels the 6d $\CN=(2, 0)$ theory. The irregular punctures require more elaborate classifications, and generally lead to non-conformal theories. But when the UV curve is a sphere, we can get a SCFT with one irregular puncture and also with or without one regular puncture \cite{Xie:2012hs, Wang:2015mra}. Theories realized in this way includes Argyres-Douglas theory \cite{Argyres:1995jj,Argyres:1995xn} and its generalizations. 

The (generalized) Argyres-Douglas theories are inherently strongly-coupled and have no weak-coupling limit. Therefore there has been no direct way of computing the superconformal indices or $S^1 \times S^3$ partition functions. Recently, a progress is made in \cite{Buican:2015ina}, where they obtained an ansatz for the TQFT description of the Schur index for some of the Argyres-Douglas type theories. They were able to verify their result against S-duality \cite{Buican:2014hfa} and also by studying dimensional reduction to 3d \cite{Buican:2015hsa}. Their result has been recently extended to the Macdonald index \cite{Buican:2015tda}. 
The result of \cite{Buican:2015ina} agrees with the general prediction made in \cite{Beem:2013sza}, and studied further in \cite{Beem:2014rza, Lemos:2014lua}, that for any $\CN=2$ $d=4$ SCFT, there is a protected sector with infinite dimensional chiral algebra acting on it. This implies that the Schur indices have to be given by the vacuum character of the corresponding chiral algebra. 

Another progress is made in \cite{Cordova:2015nma}. They observed that the trace of (inverse) monodromy operator appears in the BPS degeneracy counting \cite{Cecotti:2010fi} agrees with the Schur index (of a non-conformal theory). From this observation, they were able to predict the Schur indices of various generalized Argyres-Douglas theories and conjectured that they are given by the vacuum character of certain non-unitary $W$-minimal models. This agrees with the result of \cite{Beem:2013sza}, once applied to the Argyres-Douglas theories with no flavor symmetry, yield chiral algebra given by Virasoro/W-algebra with central charges of the minimal model series.  

In this paper, we propose the Schur, Hall-Littlewood (HL) and Macdonald limit of the superconformal index for a class $\CS$ theory  containing certain irregular puncture called (a subset of) type $I_{k, N}$.  We conjecture the wave function for irregular puncture in these limits, which enables us to use the TQFT description to compute the indices of AD theories. Our strategy is very similar to \cite{Buican:2015ina, Buican:2015tda}, but we are mainly interested in the theories with no flavor symmetry. It turns out the corresponding wave functions are much simpler than the ones with flavor symmetries.  We first start with the irregular punctures that appear in the Lagrangian theories and derive their wave function as an integral transformation of regular punctures. From here, we extrapolate the expression to find a well-behaved wave function corresponding to other irregular punctures.  For the Schur index, we are able to find the wave function for arbitrary coprime $k, N$. For the HL index and Macdonald index, we find the wave functions for the $k=2$ cases only. 

Curiously, we find that the indices for non-conformal theories can also be written in terms of the TQFT. This is the index of the theory at the UV fixed point (zero coupling) with Gauss law constraint. Even though the theory is non-conformal for non-zero gauge coupling, the UV index is nevertheless well-defined. The TQFT description for the ``superconformal index" of a non-conformal $\CN=2$ theory enables us to write the index for the conformal 4d $\CN=1$ class $\CS$ theory \cite{Benini:2009mz, Bah:2012dg, Beem:2012yn, Xie:2013gma}. 

The outline of this paper is as follows. In section \ref{sec:Schur}, we study Schur index for the theories with $I_{k, N}$ punctures. We are able to give a expression when $k$ and $N$ are coprime, and it agrees with other proposals. In section \ref{sec:HL}, we study Hall-Littlewood index with $I_{2, N}$ punctures. We obtain the wave function for $I_{2, N}$ and compare with the direct computation of the index from the 3d mirror theory. In section \ref{sec:Mac}, we conjecture a closed-form formula for $I_{2, N}$ with $N$ odd and perform a number of consistency checks. In section \ref{sec:N1}, we compute mixed Schur limit of the index for $\CN=1$ class $\CS$ theories using the result of \ref{sec:Schur}. 

%While preparing this paper, very interesting article \cite{Buican:2015tda} appeared. It provides a closed-form expression for the wave function of $I_{2, N}$ puncture in the Macdonald limit when $N$ is even and perform various checks. In the current paper, we mainly discuss the case with $N$ odd. We also provide expressions for the Schur indices of higher-rank theories. We made some consistency check between their results and ours. 

\section{Schur index} \label{sec:Schur}
Superconformal index of an $\CN=2$ $d=4$ SCFT is defined as
\be
 I(x_i; p, q, t) = \Tr (-1)^F p^{j_2 + j_1 - r} q^{j_2 - j_1 - r} t^{R+r} \prod_i x_i^{F_i} \ , 
\ee
where $j_1, j_2$ are the Cartans of the Lorentz group $SU(2)_1 \times SU(2)_2$ and $R, r$ are the generators of the $SU(2)_R$ and $U(1)_r$ symmetry respectively. $F_i$ are the generators of the flavor symmetries. The trace is taken over the $\frac{1}{8}$-BPS states that are annihilated by a supercharge $Q$. The index can be simplified by taking certain limits \cite{Gadde:2011uv}. When $p\to0$, it is called the Macdonald index and gets contributions from $\frac{1}{4}$-BPS states. The Macdonald index can be further simplified to the Schur limit upon taking $q=t$ limit. The other limit is to take $q \to 0$, and this is called the Hall-Littlewood (HL) index. 

For any class $\CS$ theories coming from 6d $(2, 0)$ theory of type $\G$ wrapped on a Riemann surface $\CC_{g, n}$, the superconformal index can be written in terms of a correlation function of a topological field theory
\be
 I(\vec{a}_i; p, q, t) = \sum_{\vec \lambda} C_{\vec \lambda}^{2g-2+n} \prod_{i=1}^n \psi^{(i)}_{\vec \lambda} (\vec{a}_i; p, q, t) \ , 
\ee
where the sum is over all irreducible representations of $\G$ and $g$ and $n$ is the genus and the number of punctures respectively. The function $C_{\vec{\lambda}}(p, q, t)$ is sometimes called the structure constant of the theory, and $\psi_{\vec{\lambda}}(\vec{a}; p, q, t)$ is called the wave function we assign to each puncture. 

The Schur index is obtained by specialization $p=0, q=t$. In this limit, the structure constant is fixed to be 
\be
C_{\vec \lambda}^{-1} = \psi^\varnothing_{\vec \lambda} (\vec{a} = q^{\vec \rho}; q) = \frac{\chi_{\vec \lambda}(q^{\vec \rho})}{\prod_{i=1}^r (q^{d_i}; q)} \ ,
\ee
where $\vec \rho$ is the Weyl vector of $\G=ADE$ and $d_i$ are the degrees of the Casmirs. We use the short-hand notation $\vec{z}^{\vec{a}} \equiv \prod_i z_i^{a_i}$ and $q^{\vec{\rho}} \equiv \prod_i q^{\rho_i}$. 
Here the $q$-Pochhammer symbol is defined as $(z; q) \equiv \prod_{i=0}^\infty (1-z q^i)$. 
For the case of a full regular puncture, the corresponding wave function is given by
\be
 \psi_{\vec \lambda} (\vec{z}; q) = \textrm{PE}\left(\frac{q}{1-q} \chi_{\textrm{adj}}(\vec{z}) \right) \chi_{\vec \lambda}(\vec{z}) = \frac{1}{(q; q)^r \prod_{\a \in \Delta} (q \vec{z}^{\a}; q)} \chi_{\vec{\lambda}} (\vec{z}) \ , 
\ee
where $\Delta$ is the set of all roots of $\G$ and $r=\textrm{rank}(\G)$. The Plethystic exponent is defined as
\be
 \textrm{PE} \left[n z \right] = \frac{1}{(1-z)^n} \ , 
\ee
where $n$ is an integer and $z$ is some fugacity. 
The wave functions are orthonormal under the vector multiplet measure
\be
 \oint [d\vec{z}] I_{\textrm{vec}} (\vec{z}) \psi_{\vec{\lambda}} (\vec{z}; q) \psi_{\vec{\mu}} (\vec{z}; q) = \delta_{\vec{\lambda}, \vec{\mu}} \ , 
\ee
where $[d\vec{z}] = \prod_{i=1}^{N-1} \frac{dz_i}{2\pi i z_i} \Delta(\vec{z})$ and $\Delta(\vec{z}) = \prod_{\a \in \Delta} (1-z^\a)$ is the Haar measure. The vector multiplet index is given by
\be
 I_{\textrm{vec}} (\vec{z}) = \PE \left( \frac{-2q}{1-q} \chi_{\textrm{adj}}(\vec{z}) \right) = (q; q)^{2r} \prod_{\a \in \Delta} (q \vec{z}^\a; q)^2 \ . 
\ee

In addition to the regular punctures, we also have irregular punctures that has higher order singularities. Let us specialize to $\G=A_{k-1}$ and denote the local singularity around the puncture at $z=\infty$ is given as 
\be
 x^k = z^N (dz)^k \ , 
\ee
or equivalently around $z'=0$ via $z'=1/z$ as 
\be
 x^k = \frac{1}{z'^{N+2k}} (dz')^k \ , 
\ee
to be $I_{k, N}$, following the notation of \cite{Xie:2012hs,Xie:2013jc}. One can further deform the singularity by adding less singular terms. They serve as deformation parameters of the theory. The regular punctures correspond to $N=-k$. For the case of $k=2$ and $N=2n$ even, the wave function is written by \cite{Buican:2015ina} as
\be 
 \psi^{I_{2, 2n}}_{\lambda} (a, q) = \frac{q^{(n+1)\frac{\lambda}{2}\left(\frac{\lambda}{2}+1 \right)}}{(q; q)} \Tr_{R_\lambda} ( a^{2j_3} q^{-(n+1) j_3^2 })  \ ,  
\ee
where $R_\lambda$ is the spin-$\frac{\lambda}{2}$ representation of $SU(2)$, and $j_3 = -\frac{\lambda}{2}, -\frac{\lambda}{2}+1, \cdots, \frac{\lambda}{2}$. 

Our goal in this section is to find a wave function for the punctures of type $I_{k, N}$. For example, we conjecture the wave function for the $I_{2, 2n-1}$ puncture to be 
\be
\psi^{I_{2, 2n-1}}_{\lambda} (q) =
\begin{cases}
  (-1)^{\frac{\lambda}{2}} q^{\frac{\lambda}{2}\left( \frac{\lambda}{2}+1 \right)(n+\half)} & \lambda \textrm{ even} , \\
  0 & \textrm{otherwise}  ,
\end{cases} 
\ee 
as we will provide evidences in this section. 

\subsection{Wave function for the puncture of type $I_{k, N}$}
\paragraph{Puncture of type $I_{N, -N+1}$} 
Let us consider $SU(N)$ theory with $N$ flavors. This theory can be realized by a 3-punctured sphere with one maximal, one minimal and one irregular of type $I_{N, -N+1}$. If we write the index of this theory in terms of a TQFT, we get
\be
 I (\vec{a}, x) = \sum_{\vec{\lambda}} C_{\vec{\lambda}} \psi_{\vec\lambda} (\vec{a}) \psi^{\star}_{\vec{\lambda}}(x) \psi^{I_{N, -N+1}}_{\vec\lambda} \ , 
\ee
where $\psi^{\star}_{\vec{\lambda}}(x)$ and $\psi^{I_{N, -N+1}}_{\vec\lambda} $ denote the wave function corresponding to the minimal and $I_{N, -N+1}$ puncture respectively. 
Since it is the same as gauging one of the flavors of bifundamental hypermutliplets, it can be also written as
\be
 I(\vec{a}, x) = \oint [d\vec{z}] I_{\textrm{vec}}(\vec{z}) \sum_{\vec{\lambda}} C_{\vec{\lambda}} \psi_{\vec\lambda} (\vec{a}) \psi^{\star}_{\vec{\lambda}}(x) \psi_{\vec\lambda} (\vec{z}) \ .
\ee
This means that we can simply write the wave function corresponding to the irregular puncture $I_{N, -N+1}$ as\footnote{The author would like to thank Yuji Tachikawa for the discussions lead to this observation.}
\be \label{eq:WaveftnGW}
\psi^{I_{N, -N+1}}_{\vec\lambda} (q) = \oint [d\vec{z}] I_{\textrm{vec}}(\vec{z}) \psi_{\vec\lambda} (\vec{z}) 
= \oint [d\vec{z}] \PE \left[ -\frac{q}{1-q} \chi_{\textrm{adj}} (\vec{z}) \right] \chi_{\vec{\lambda}}(\vec{z})  \ . 
\ee
This integral can be written in a more illuminating form by using the Jacobi triple identity
\be \label{eq:Jacobi}
 (q; q) (y^{-1}; q) (q y; q) = \sum_{m \in \IZ} (-1)^m q^{\half m(m+1)} y^m \ . 
\ee
We get
\begin{align}
\psi_{\vec{\lambda}}^{I_{N, -N+1}} (q) &= \frac{1}{|\CW| (q; q)^{\half (N-1)(N-2)}}  \\ 
  &~~~ \times \sum_{n_\a \in \IZ} (-1)^{\sum_\a n_\a} q^{\sum_\a \half n_\a (n_\a +1)} \oint d\vec{z} \prod_{\a \in \Delta^+} (1-\vec{z}^\a) \vec{z}^{\sum_\a n_\a \a} \chi_{\vec{\lambda}}(\vec z) \nn \ ,
\end{align}
where $\Delta_+$ is the set of positive roots of $SU(N)$ and $\CW$ is the Weyl group. Here the index $\a$ runs from $\a = 1, \cdots |\Delta_+|$ and the integral measure is given by $d\vec{z} = \prod_{i=1}^{N-1} \frac{dz_i}{2\pi i z_i}$. The last integral can be rewritten upon applying the Weyl character formula as
\be
 \sum_{w \in \CW} \e(w) \oint d\vec{z} z^{-\rho + w(\vec{\lambda} + \rho) - \sum_{\a} n_\a \a }
  = \sum_{w \in \CW} \e(w) \delta^{(N-1)}_{w \cdot \vec {\lambda} = n_\a \vec{\a}} \ , 
\ee
where $\rho = \half \sum_{\a \in \Delta_+} \a$ is the Weyl vector, $\e(w)$ is the signature of $w$ which is the same as the determinant of $w$ and $w \cdot \vec{\lambda} \equiv w(\vec{\lambda}+\rho)-\rho$ is the shifted Weyl reflection. 

For the case of $N=2$, it is simple to evaluate the above integral. We get
\begin{align}
 \psi_\lambda^{I_{2, -1}}(q) &= \half \sum_{n \in \IZ} (-1)^n q^{\half n(n+1)} \oint \frac{dz}{2\pi i z} (z^{\lambda - 2n} -z^{-\lambda-2n-2} ) \nn \\
  &=  \begin{cases} (-1)^{\frac{\lambda}{2}} q^{\half \frac{\lambda}{2} (\frac{\lambda}{2}+1)} & \mbox{$\lambda$ even}, ~~ \\ 
0 & \mbox{$\lambda$ odd}.
 \end{cases} 
\end{align}
For $N=3$, we find the expression as below:
\be
 \psi^{I_{3, -2}}_{(\lambda_1, \lambda_2)}(q) = 
 \begin{cases}
  q^{k(k+1)+\ell(\ell+1)+k \ell} & \mbox{if $\lambda_1 = 3k, \lambda_2 = 3\ell$} , \\
  -q^{k^2+\ell^2-1+(k-1)(\ell-1)} & \mbox{if $\lambda_1 = 3k-2,  \lambda_2 = 3 \ell-2 $} , \\
  0 & \mbox{otherwise} , 
 \end{cases}
\ee
where $k, \ell \in \IZ_{\ge 0}$. 
We do not have an analytic proof of the formula, but we have checked this expression up to $(\lambda_1, \lambda_2)=(12, 12)$. 

This wave function is an analog of Gaiotto-Whittaker vector \cite{Gaiotto:2009ma, Taki:2009zd, Keller:2011ek} in the AGT correspondence \cite{Alday:2009aq, Wyllard:2009hg}, which realize pure YM theory when we have two punctures of this type. In our case, we expect the two point function of Gaiotto-Whittaker state of the $q$-deformed Yang-Mills gives us the `Schur index' of the pure YM theory\footnote{For a non-conformal theory, the superconformal index really means that of the free UV fixed point with the Gauss-law constraint.}
\be \label{eq:YMtqft}
 I_{YM}(q) = \sum_{\vec\lambda} \psi^{I_{N, -N+1}}_{\vec\lambda}(q) \psi^{I_{N, -N+1}}_{\vec\lambda}(q) \ . 
\ee
We will prove this relation for the case of $SU(2)$ in section \ref{subsec:su2sym}.

\paragraph{Type $I_{k, N}$ with $(k, N) = 1$}
Now, we conjecture that the wave function for the irregular puncture of type $I_{k, N}$ when $k$ and $N$ are relative primes, is simply given by rescaling the $q$ parameter of the wave function for the Gaiotto-Whittaker state \eqref{eq:WaveftnGW}. More precisely, we find
\be \label{eq:Ikn}
 \psi^{I_{k, N}}_{\vec \lambda} (q) = \psi^{I_{k, 1-k}}_{\vec \lambda} (q^{N+k}) \ . 
\ee
This can be thought of as an analog of the coherent state considered by \cite{Bonelli:2011aa,Gaiotto:2012sf,Kanno:2013vi} in the context of AGT correspondence. 
We give a number of evidences for \eqref{eq:Ikn} in later sections. It would be desirable to give a direct proof of this proposal.%\footnote{Here is one heuristic argument why this simple rescaling of $q \to q^n$ makes sense. When the index is considered as a partition function on $S^1 \times S^3$, the fugacity $q$ is given by the ratio of the size of $S^1$ and $S^3$ as $q=\exp(-R_{S^1}/R_{S^3})$. From this perspective, $q \to q^n$ means taking $R_{S^1}$ to $n R_{S^1}$. The irregular puncture behave as (for $A_1$ theory) $\Phi^{(n)}(z) \sim z^{-n-1/2} dz$ near the singularity. Here $n=1$ for the Gaiotto state (or $I_{2, -1}$). The holonomy $\a_n \sim \oint \Phi^{(n)} (z) $ satisfies $\a_n = \a_1/n$. In the $q$-YM, the holonomy is valued in $S^1$, which makes $\a_n$ to have periodicity $2\pi n R_{S^1}$, which effectively makes $R_{S^1}$ around irregular singularity to be $n$ times larger.}  

\paragraph{Puncture of type $I_{k, kn}$}
We can also consider a irregular puncture of type $I_{k, N}$ with $N$ being a multiple of $k$. In this case we have $k-1$ mass parameters associated to the $U(1)^{k-1}$ flavor symmetry. As before, let us first consider a Lagrangian example. Consider $SU(N)$ gauge theory with $2N-1$ fundamentals. It is realized by a sphere with a maximal, minimal and $I_{N, 0}$ type irregular punctures. Therefore, we write the wave function for the $I_{N, 0}$ puncture as
\be
\begin{split}
 \psi^{I_{N, 0}}_{\vec \lambda} (\vec{a}; q) &= \oint [d\vec{z}] I_{\textrm{vec}}(\vec{z}) I_{\textrm{hyp}}(\vec{z}, \vec{a}) \psi_{\vec\lambda}(\vec{z}) \\
 &= \oint [d\vec{z}] \PE \left[ -\frac{q}{1-q} \chi_{\textrm adj}(\vec{z}) \right] \PE \left[ \frac{q^{\half}}{1-q}\sum_{i=1}^N \sum_{m=1}^{N-1} (z_i a_m + z_i^{-1} a_m^{-1} ) \right] \chi_{\vec\lambda}(\vec{z}) ~~\\
 &= \oint [d\vec{z}] (q; q)^r \prod_{i \neq j} (q z_i/z_j; q) \prod_{i=1}^{N} \prod_{m=1}^{N-1} \frac{1}{(q^{\half} (z_i a_m)^\pm; q)} \chi_{\vec\lambda}(\vec{z}) \ ,
 \end{split}
\ee
where we impose $\prod_i z_i=1$ and $\prod_m a_m =1$. Here $\pm$ means taking product of each sign. 

For the case of $N=2$, the wave function is given by \cite{Buican:2015ina} as
\be \label{eq:I22n}
 \psi^{I_{2, 2n}}_{\lambda} (a, q) = \frac{q^{(n+1)\frac{\lambda}{2} (\frac{\lambda}{2}+1)}}{(q;q)} \Tr_{R_\lambda} ( a^{2 j_3} q^{-(n+1) j_3^2 })  \ , 
\ee
where the trace is over the spin-$\lambda/2$ representation $R_\lambda$. 
We find that the wave function for the irregular puncture of type $I_{2, 2n}$ can be written almost as a simple rescaling of the $q$ as
\be
 \psi^{I_{2, 2n}}_{\lambda} (a;q) = \frac{(q^{n+1}; q^{n+1})}{(q;q)} \psi^{I_{2, 0}}_{\lambda} (a; q^{n+1}) \ . 
\ee
We have checked this expression indeed gives us the same wave function as \eqref{eq:I22n} found in \cite{Buican:2015ina} to high orders in $q$.

We were not able to find a prescription for $N\ge3$. It may be possible to find a correct prescription by using isomorphism of $(A_2, A_2) = (A_1, D_4)$, which has the chiral algebra $\widehat{su}(3)_{-\frac{3}{2}}$.

\subsection{Examples}
\subsubsection{Lagrangian theories} \label{subsec:su2sym}

\paragraph{$SU(2)$ SYM}
Pure $SU(2)$ YM theory can be realized on a sphere with 2 irregular punctures with $x^2 \propto (dz')^2 /z'^3$, which means $I_{2, -1}$. From the TQFT, we get
\be
 I_{\textrm{TQFT}(-1, -1)} (q) = \sum_{\lambda} \psi_\lambda^{I_{2, -1}} \psi_\lambda^{I_{2, -1}} = \sum_{j \in \IZ_{\ge 0}} q^{j(j+1)} \ . 
\ee
It agrees with the integral expression obtained from blindly applying the index formula for the vector multiplets and then integrating over the gauge group
\be
 I_{\textrm{SYM}} (q) = (q; q)^2 \oint \frac{dz}{2\pi i z} \Delta(z) (q z^{\pm 2}; q)^2 = \sum_{m \in \IZ_{\ge 0}} q^{m(m+1)} \ ,  
\ee
where $\Delta(z) = \half (1-z^{\pm 2})$ is the Haar measure of $SU(2)$. This integral can be easily evaluated by using the Jacobi triple product identity \eqref{eq:Jacobi}. It can be considered as the index at the UV fixed point with Gauss law constraint. 

\paragraph{$SU(2)$ with $N_f=1$}
The $SU(2)$ gauge theory with 1 flavor can be realized by a sphere with 2 irregular punctures $I_{2, -1}$ and $I_{2, 0}$. From the TQFT, we get
\be
\begin{split}
 I_{\textrm{TQFT}(-1, 0)} (q, a) &= \sum_{\lambda} \psi_\lambda^{I_{2, -1}} \psi_\lambda^{I_{2, 0}}(a) \\
 &= 
 1+q+\left(-a^2-\frac{1}{a^2}+2\right) q^2+\left(-a^2-\frac{1}{a^2}+2\right) q^3  \\
 & \quad + \left(-2 a^2-\frac{2}{a^2}+4\right) q^4 
   +\left(a^4+\frac{1}{a^4}-3 a^2-\frac{3}{a^2}+5\right) q^5+O\left(q^6\right) \ ,
\end{split}
\ee
which agrees with the one computed from the integral
\be
 I_{N_f=1} = (q;q)^2 \oint \frac{dz}{2\pi i z} \Delta(z) \frac{(q z^{\pm 2}; q)^2}{(q^\half z^{\pm} a^{\pm}; q)} \ . 
\ee

\paragraph{$SU(2)$ with $N_f=2$}
The $SU(2)$ gauge theory with 2 flavors can be realized by a sphere with 2 irregular punctures $I_{2, 0}$. This gives us
\be
 I_{\textrm{TQFT}(0, 0)} &=& \sum_{\lambda} \psi_\lambda^{I_{2, 0}}(a) \psi_\lambda^{I_{2, 0}}(b) \\
  &=& 1 + q \left(\chi^{SO(4)}_{[1, 0]} + \chi^{SO(4)}_{[0, 1]} + 2 \chi^{SO(4)}_{[0, 0]} \right) + q^2 \left(\chi^{SO(4)}_{[2, 0]} + \chi^{SO(4)}_{[0, 2]} + 2 \chi^{SO(4)}_{[1, 0]}+ 2 \chi^{SO(4)}_{[0, 1]}+ 3 \chi^{SO(4)}_{[0, 0]} \right) \nn \\
 &{ }&\quad + q^3 \left( \chi^{SO(4)}_{[3, 0]} + \chi^{SO(4)}_{[0, 3]} + 2 \chi^{SO(4)}_{[2, 0]} + 2 \chi^{SO(4)}_{[0, 2]} + 4 \chi^{SO(4)}_{[1, 0]} + 4 \chi^{SO(4)}_{[0, 1]} + 6 \chi^{SO(4)}_{[0, 0]} \right) + O(q^4) \ , \nn
\ee
which agrees with the computation from the integral formula. Here we used the Dynkin label to denote the characters. 

We can also realize the same theory via 3-punctured sphere with 2 regular punctures and one $I_{2, -1}$ puncture. This gives
\be
 I_{\textrm{TQFT}(R, R, -1)} (x, y; q) = \sum_\lambda C_\lambda \psi_\lambda (x) \psi_\lambda (y) \psi_\lambda^{I_{2, -1}} \ , 
\ee
which gives the same answer upon matching the fugacities via $a \to \sqrt{xy}$ and $b \to \sqrt{x/y}$. 

\paragraph{$SU(2)$ with $N_f=3$}
The $SU(2)$ gauge theory with 3 flavors can be realized by a sphere with 2 regular punctures and 1 irregular puncture $I_{2, 0}$. This gives us the index to be 
\be
\begin{split}
I_{\textrm{TQFT}(R, R, 0)} &= \sum_\lambda C_\lambda \psi_\lambda (a) \psi_\lambda (b) \psi_\lambda^{I_{2, 0}}(c) \\ 
 &= 1 + q \chi^{SO(6)}_{[0, 1, 1]}  + q^2 (\chi^{SO(6)}_{[0, 2, 2]}+\chi^{SO(6)}_{[0, 1, 1]} +\chi^{SO(6)}_{[0, 0, 0]}) \\
 &\quad + q^3 (\chi^{SO(6)}_{[0, 3, 3]}+\chi^{SO(6)}_{[0, 2, 2]}+\chi^{SO(6)}_{[1, 2, 0]}+\chi^{SO(6)}_{[1, 0, 2]}+\chi^{SO(6)}_{[0, 1, 1]}+\chi^{SO(6)}_{[0, 0, 0]}) + O(q^4)\ . 
\end{split}
\ee
This agrees with the integral formula
\be
 I_{N_f=3} (a_i; q) = (q; q)^2 \oint \frac{dz}{2\pi i z} \Delta(z) \frac{(q z^{\pm 2}; q)^2}{\prod_{i=1}^3 (q^\half z^\pm a_i^\pm ; q) }\ ,
\ee
upon identifying $a_1 = bc, a_2 = b/c, a_3 = a$. 

\paragraph{$SU(3)$ SYM} 
The Schur index of the $SU(3)$ pure YM can be written as
\be \begin{split}
 I_{SU(3)} &= \frac{1}{3!} (q; q)^4 \oint \frac{dz_1}{2\pi i z_1} \frac{dz_2}{2\pi i z_2} \prod_{\a \in \Delta_+} (1-z^{\pm \a}) (q z^{\pm\a}; q)^2 \\
 &= \frac{1}{6 (q;q)^2}  \sum_{n_{1,2,3}, m_{1, 2, 3} \in \IZ} (-1)^{n_1+m_1} q^{\half \sum_i \left(n_i(n_i+1) + m_i (m_i+1) \right) } \delta_{n_1+n_2, m_1+m_2} \delta_{n_1+n_3, m_1+m_3} \ . 
 \end{split}
\ee
We verified this to be the same as the one given by \eqref{eq:YMtqft} with $N=3$ to high orders in $q$. The first few terms are
\be
I_{SU(3)} = 1+q^2+2 q^4+2 q^8+q^{10}+2 q^{12}+q^{16}+2 q^{18}+2 q^{20}+ O(q^{24}) \ . 
\ee

\subsubsection{Argyres-Douglas theories}
\paragraph{$(A_1, A_{N-1})$ theories}

The Argyres-Douglas theories of type $(A_1, A_{N-1})$ can be realized by a sphere with single irregular puncture of type $I_{2, N}$ \cite{Xie:2012hs,Xie:2013jc}. When $N=2n+1$ is odd, we have
\be \label{eq:A1A2nSchur}
 I_{(A_1, A_{2n})}(q) = \sum_{\lambda} C_\lambda^{-1} \psi_\lambda^{I_{2, 2n+1}} = \sum_{j \in \IZ_{\ge 0}} \frac{[2j+1]_q}{(q^2; q)} (-1)^j q^{j(j+1)(n+\frac{3}{2})} \ , 
\ee
where 
\be
C_\lambda^{-1} = \frac{[\textrm{dim}\lambda]_q}{(q^2; q)}  \ , \qquad [n]_q = \frac{q^{\frac{n}{2}} - q^{-\frac{n}{2}}}{q^{\half} - q^{-\half}} \ . 
\ee

The vacuum character of the Virasoro minimal model $(r, s)$ is given by 
\be
 \chi_0^{(r, s)} (q) = \frac{q^{-\frac{1}{4} \frac{(r-s)^2}{r s}}}{(q;q)} \sum_{\ell \in \IZ} \left( q^{\frac{(2r s \ell + r - s)^2}{4 r s}} - q^{\frac{(2r s \ell + r + s)^2}{4 r s}} \right) \ . 
\ee
We have checked this expression agrees with the vacuum character of the Virasoro minimal model $(2, 2n+3)$ to high orders in $q$, 
which is consistent with the relation given in \cite{Beem:2013sza}, and also computation based on the conjectural relation from the BPS degeneracy \cite{Cordova:2015nma}. We also find that when $n=0$ or $N=1$, the index becomes $1$, which agrees with the expectation that there is no massless degrees of freedom and vanishing central charges \cite{Xie:2013jc} for the $(A_1, A_0)$ theory. 

When $N=2n$ is even, we get \cite{Buican:2015ina}
\be
 I_{(A_1, A_{2n-1})}(x; q) = \sum_{\lambda} C_\lambda^{-1} \psi_{\lambda}^{I_{2, 2n}}(x) = \sum_{\lambda \ge 0} \frac{[\lambda+1]_q}{(q^2; q)}  \Tr_{R_\lambda} \left( x^{2j_3} q^{(n+1)\left( \frac{\lambda}{2}( \frac{\lambda}{2}+1 )-j_3^2 \right)} \right)  , 
\ee
which agrees with the vacuum character of $\widehat{su}(2)_{-\frac{4}{3}}$ when $n=2$. 

\paragraph{$(A_1, D_{N+2})$ theories}
The AD theory of type $(A_1, D_{N+2})$ can be realized by a sphere with one $I_{2, N}$ type irregular puncture and a regular puncture. 
When $N=2n-1$, the index can be written as 
\be
I_{(A_1, D_{2n+1})} &=& \sum_{\lambda} \psi_\lambda^{I_{2, 2n-1}} \psi_\lambda (x) 
= \frac{1}{(q x^{\pm 2, 0}; q)} \sum_{m=0}^\infty (-1)^m q^{\frac{m(m+1)}{2} (2n+1)} \ch_{R_{2m}}(x)  . 
\ee
This expression exactly agrees with the vacuum character of the affine Lie algebra $\widehat{su}(2)_{-\frac{4n}{2n+1}}$. When $N=1$, we find the index of $(A_1, A_3)$ agrees with that of $(A_1, D_3)$ 
\be
 I_{(A_1, A_3)} = \sum_\lambda C_\lambda^{-1} \psi_\lambda^{I_{2, 4}}(x) = \sum_\lambda \psi_\lambda^{I_{2, 1}} \psi_\lambda (a) = I_{(A_1, D_3)} \ , 
\ee
up on identifying $x=a^2$. 

When $N=2n$, the index can be written as
\be
I_{(A_1, D_{2n+2})} &=& \sum_{\lambda} \psi_\lambda^{I_{2, 2n}}(x) \psi_\lambda (y) \ . 
\ee
When $n=1$, this agrees with the vacuum character of $\widehat{su}(3)_{-\frac{3}{2}}$ \cite{Buican:2015ina}. 

\paragraph{$(A_{k-1}, A_{N-1})$ theories with $(k, N)=1$}
When $k$ and $N$ are coprime, we conjecture that the Schur indices for the Argyres-Douglas theories of type $(A_{k-1}, A_{N-1})$ as
\be
 I_{(A_{k-1}, A_{N-1})}(q) = \frac{1}{\prod_{i=2}^{k} (q^i; q)} \sum_{\vec \lambda} \chi_{\vec \lambda}(q^\rho) \psi^{I_{k,N}}_{\vec \lambda}(q) \ . 
\ee
It was conjectured by \cite{Cordova:2015nma} that for the coprime $k, N$, the Schur index of $(A_{k-1}, A_{N-1})$ theory is given by the vacuum character of the $(k, k+N)$ $W_k$-minimal model, which is given by \cite{andrews1999a}
\be
\chi_0^{W(k, k+N)}(q) = \left( \frac{(q^{k+N}; q^{k+N})}{(q;q)} \right)^{k-1} \prod_{a=1}^{k-1} (q^{N+a}; q^{k+N})^a (q^a; q^{k+N})^{k-a} \ .
\ee
We checked $I_{(A_{k-1}, A_{N-1})}(q) = \chi_0^{W(k, k+N)}(q) $ to high orders in $q$ for a number of cases.

%%%%%%%%%%%%%%%%%%%%%%%%%%%%%%%%%%%%%%%%%%%%%%%%%

\section{Hall-Littlewood index} \label{sec:HL}

\subsection{Index from the 3d mirror}
For Argyres-Douglas theories realized in class $\CS$ with a non-trivial Higgs branch\footnote{Even though the theory in the UV might not have a Higgs branch, AD points can sometimes have a quantum Higgs branch \cite{Argyres:2012fu}. }, once dimensionally reduced to 3d, their mirror theories \cite{Intriligator:1996ex} are known \cite{Xie:2012hs, 2008arXiv0806.1050B}. The Higgs branch of the dimensionally reduced 3d $\CN=4$ theory is the same as the original 4d $\CN=2$ theory, which is the same as the Coulomb branch of the 3d mirror
\be
 \CM^{4d}_{\textrm{Higgs}} = \CM^{3d}_{\textrm{Higgs}} = \CM^{3d~\textrm{mirror}}_{\textrm{Coulomb}} \ . 
\ee 
Since the Hall-Littlewood (HL) index is the same as the Hilbert series of the Higgs branch \cite{Gadde:2011uv}, it should be the same as the Hilbert series of the Coulomb branch \cite{Cremonesi:2013lqa} of the 3d mirror theory. This can be also computed using the 3d superconformal index by taking the `Coulomb limit' as discussed in \cite{Razamat:2014pta}. Therefore we can write
\be
 I^{4d}_{\textrm{HL}}(t) = I^{3d}_{\textrm{Higgs}}(t) = I^{3d~\textrm{mirror}}_{\textrm{Coulomb}}(t) \ . 
\ee
 In this section, we review the computation of HL indices from the 3d mirrors \cite{DelZotto:2014kka}. 

The Coulomb branch index for the 3d $\CN=4$ theory is defined as
\be
 I^C (t) = \Tr (-1)^F t^{E - R_H} = \Tr(-1)^F t^{E + j} \ , 
\ee
where the trace is over the states with $E = R_C$ in addition to $E - R_H-E_C - j = 0$. Here $R_H, R_C$ are the Cartans of $SO(4)_R = SU(2)_{R_H} \times SU(2)_{R_C}$ and $E$ is the scaling dimension and $j$ is the Cartan of the rotation group $SO(3)$. 
The Coulomb branch index contribution for a hypermultiplet is simply
\be
 I^C_{\textrm{hyp}} (z, m; t) = t^{\frac{1}{2} |m| } \ , 
\ee
which does not depend on any flavor fugacities. Here $m$ is the charge for the background topological $U(1)$ associated to the gauge group. 

\begin{figure}[h]
	\centering
	\includegraphics[width=2.0in]{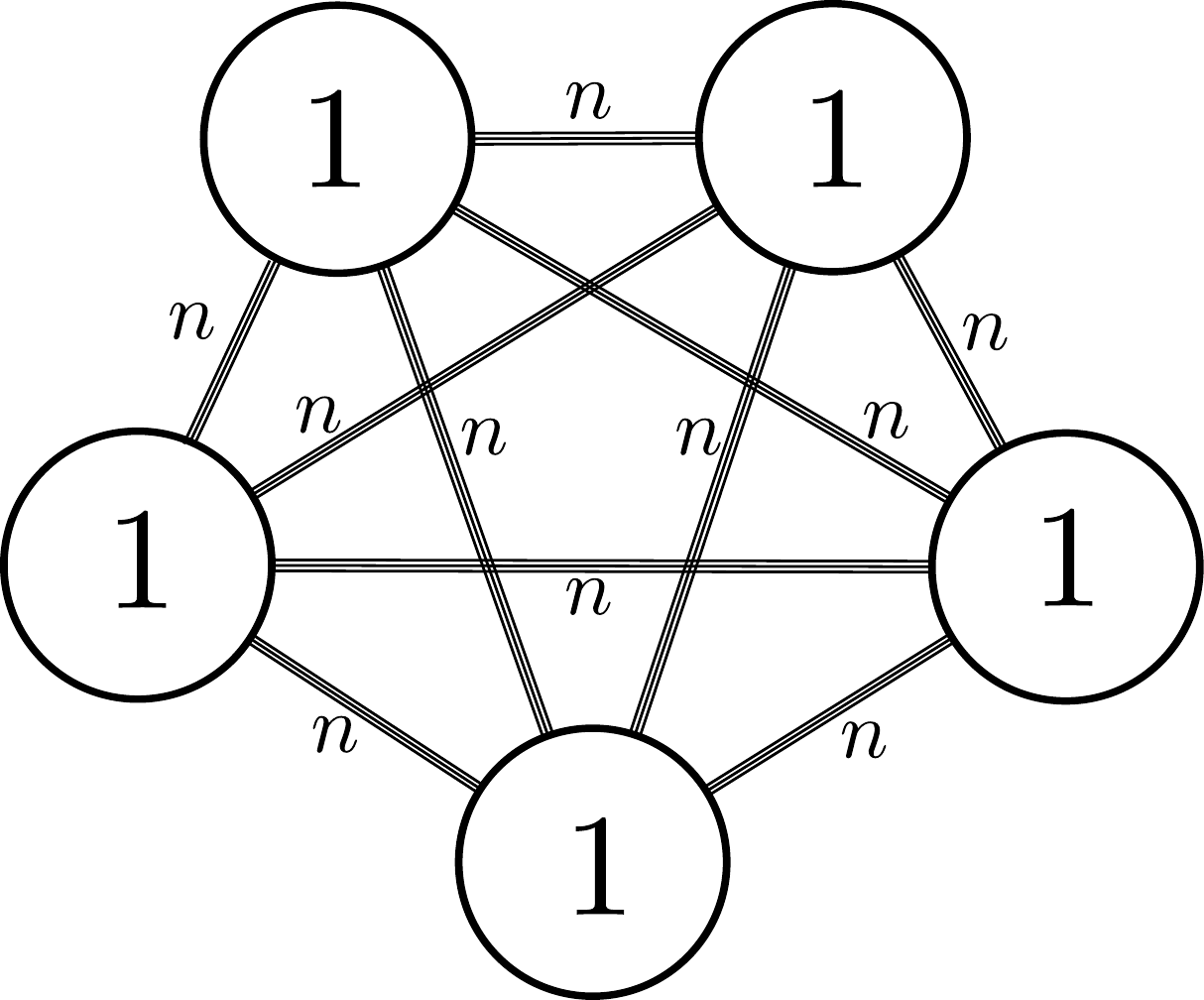}
	\caption{The 3d mirror of $(A_{k-1}, A_{kn-1})$ theory. $k$ is the number of nodes and $n$ is the number of bifundamentals between a pair of nodes. One of the $U(1)$ has to be decoupled. }
	\label{fig:MirrorQuiverAn}
\end{figure}

\paragraph{$(A_1, A_{2n-1})$ theories}
The 3d mirror theory is given by a quiver gauge theory with $k$ $U(1)$ nodes where all the nodes are connected by $n$ edges. One of the $U(1)$ node has to be ungauged as usual. See the figure \ref{fig:MirrorQuiverAn}. 
For the $U(1)$ gauge theory with $n$ fundamental hypers, the index is given as
\be \label{eq:HLidxA1}
 I^C_{(A_1, A_{2n-1})} (w; t) = \frac{1}{1-t} \sum_{m\in \IZ} w^m t^{\frac{n}{2} |m|}  \ , 
\ee
where $w$ is the fugacity for the topological $U(1)_J$ associated to the gauge group. One can sum up the formula to get
\be \label{eq:HLidxI22n}
 I^C_n (w; t) = \left(\frac{1-t^n}{1-t} \right) \frac{1}{(1-t^{\frac{n}{2}} w)(1-t^{\frac{n}{2}} w^{-1})} 
 = \left(\frac{1-t^n}{1-t} \right) \sum_{k=0}^\infty \chi^{SU(2)}_{[k]} (w) t^{k n/2} \ , 
\ee
where $\chi_{[k]}$ denotes the characters of $SU(2)$ with Dynkin label $[k]$. Even though we see the index is written in terms of characters $SU(2)$, it is wrong to say that the global symmetry is actually enhanced to $SU(2)$. Note that only when $n=2$, we have the extra conserved current coming from the monopoles \cite{Gaiotto:2008ak}. 

The Coulomb branch of $U(1)$ gauge theory with $n$ electrons or the Higgs branch of its mirror theory $\hat{A}_{n-1}$ quiver is known to be given by $\IC^2 / \IZ_n$. One can directly see that the Hilbert series of this space is the same as above. 

For $n=2$, the 3d theory is well-known $T[SU(2)]$ theory which is self-mirror. Here the $SU(2)$ enhancement of topological symmetry is evident from the mirror perspective. There $SU(2)$ symmetry is nothing but the flavor symmetry of two electrons. In this case, the 4d theory is the same as the AD theory found from $SU(2)$ gauge theory with two flavors. 

\paragraph{$(A_2, A_{3n-1})$ theories}
The 3d mirror of this theory is given by $U(1)\times U(1)$ gauge theory with $n$ bifundamentals and $n$ electrons for each $U(1)$.
The Coulomb branch index is given by
\be
 I^C_{(A_2, A_{3n-1})} (w_1, w_2; t) = \frac{1}{(1-t)^2} \sum_{m_1, m_2 \in \IZ} w_1^{m_1} w_2^{m_2} t^{\half(|m_1| + |m_2| + | m_1 - m_2 | ) n} \ , 
\ee
which can be written in terms of $SU(3)$ characters as
\be
I^C_{(A_2, A_{3n-1})} (\vec{a}; t) = \left( \frac{1-t^n}{1-t} \right)^2 \sum_{k=0}^{\infty} \chi^{SU(3)}_{[k, k]} (\vec{a}) t^{nk}
\ee
where $[k, k]$ is the Dynkin label for the $k$-th powers of adjoint representation of $SU(3)$ and the fugacities are mapped to $w_1 = a_1/a_2^2, w_2 = a_2/a_1^2$. From above, we see that the global symmetry is enhanced to $SU(3)$. 

This flavor symmetry can be understood quite easily when $n=1$. In this case, the quiver gauge theory we obtain is nothing but $\hat{A}_2$ quiver theory with $U(1)$ nodes, which is mirror to the $U(1)$ gauge theory with 3 electrons. As in the previous case, we have $SU(3)$ flavor symmetry for $n=1$, which gives us extra conserved current from the monopole operators. For $n > 1$, even though we get $SU(3)$ character representations, it does not mean that our theory has extra conserved currents. 

Curiously, we observe that the 3d mirror of $(A_2, A_2)$ theory is simply given by $\hat{A}_{2}$ quiver theory, which is mirror to $U(1)$ with $3$ electrons, which is the 3d mirror of $(A_1, A_3)$ theory. This is not a surprise, in the sense that generally in class $\CS$, there can be many different ways to realize the same 4d SCFT. See \cite{Maruyoshi:2013fwa} for a study of such examples among the AD theories.

\paragraph{$(A_1, D_{2n+2})$ theories}
This theory is obtained by putting extra $U(1)$ punctures to $(A_1, A_{2n-1})$ theory. The 3d mirror is obtained by gluing the 3d mirror theory corresponding to the minimal puncture to the 3d mirror of $(A_{k-1}, A_{kn-1})$ theory.
\begin{figure}[h] 
	\centering
	\includegraphics[width=2.0in]{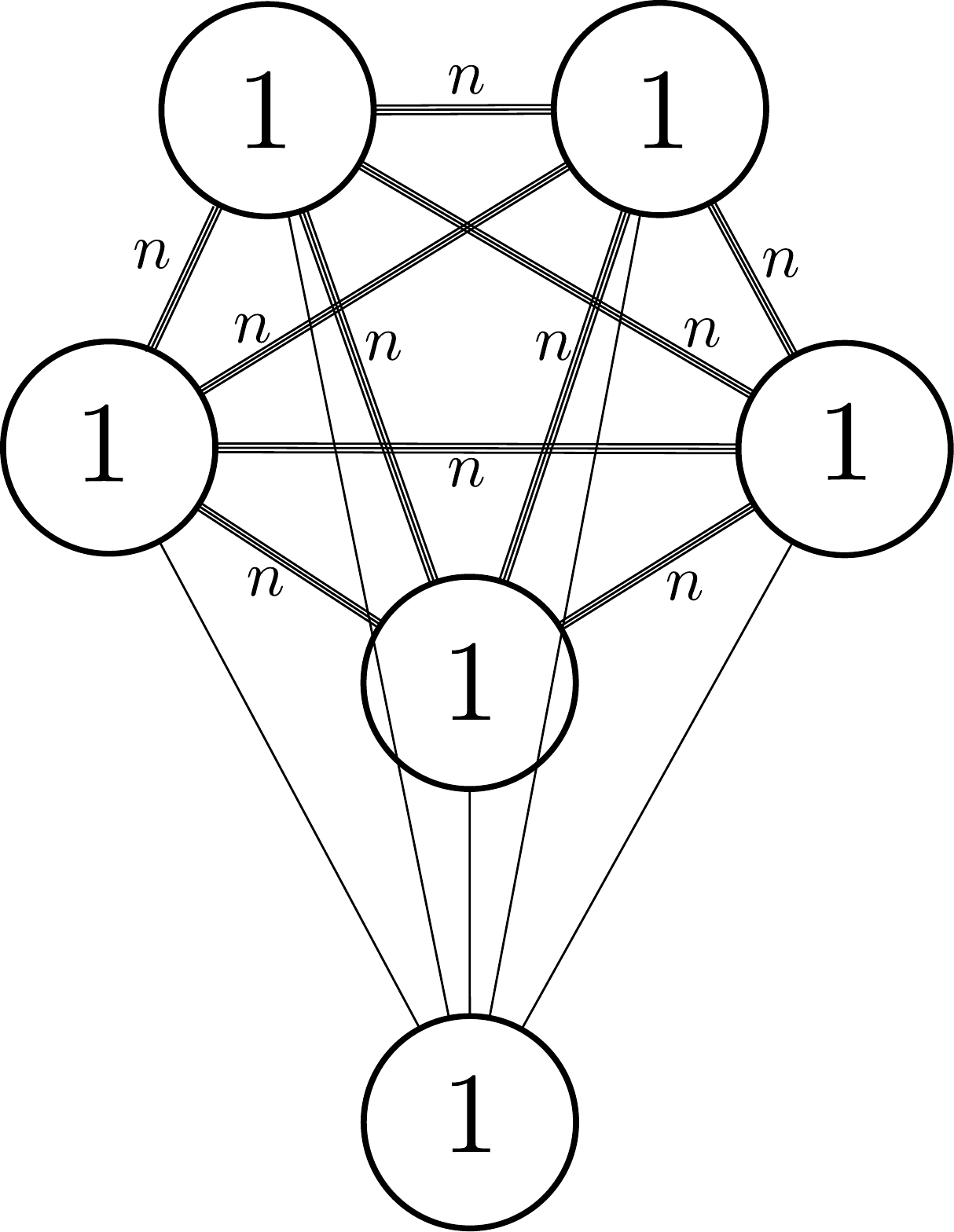}
	\caption{The 3d mirror of $(A_{k-1}, D_{kn+2})$ theory. The overall $U(1)$ has to be decoupled as before. }
	\label{fig:MirrorQuiverDn}
\end{figure}
See the figure \ref{fig:MirrorQuiverDn}. The Coulomb index for the mirror theory is given as
\be \label{eq:A1Dn3d}
 I^{C}_{(A_1, D_{2n+2})} = \frac{1}{(1-t)^2} \sum_{m_1, m_2} w_1^{m_1} w_2^{m_2} t^{\half (n |m_1 - m_2| + |m_1| + |m_2 | )} \ . 
\ee
When $n=1$, the above sum can be written as
\be
 I^{C}_{(A_1, D_4)} = \sum_k \chi^{SU(3)}_{[k, k]} (\vec{a}) t^k \ , 
\ee
which is the same as that of $(A_2, A_2)$ theory as expected.

\subsection{Wave function for $I_{2, N}$} 
\paragraph{$N=2n$ even}
We want to write the index \eqref{eq:HLidxA1} using the TQFT as
\be \label{eq:HLtqft}
 I_{(A_1, A_{2n-1})} (a) = \sum_{\lambda} C_{\lambda}^{-1} \psi_\lambda^{I_{2, 2n}}(a) \ , 
\ee
with 
\be
 C_\lambda = \frac{\prod_{i=1}^r (1-t^{d_i})}{P^{HL}_\lambda (t^{\vec{\rho}})} \ , 
\ee
where $d_i$ are the degrees of the Casmirs and $\vec\rho$ is the Weyl vector of $\G$. For $\G=SU(N)$, they are given by $d_i=2, 3, \cdots, N$. Here the function $P^{HL}_\lambda (a)$ is the Hall-Littlewood polynomial labelled by a Dynkin label $\vec{\lambda}$ in general, which we normalize so that
\be
 \langle P^{HL}_{\vec\lambda}(z), P^{HL}_{\vec\mu}(z) \rangle = \int [dz] \Delta(z) \PE \left[ t \chi_{\textrm{adj}} (\vec{z}) \right] P^{HL}_{\vec\lambda}(z) P^{HL}_{\vec\mu}(z) = \delta_{\lambda \mu} \ . 
\ee
For the case of $\G=SU(2)$, we get
\be
 P^{HL}_\lambda (z) = 
 \begin{cases}
\sqrt{1-t^2} & \mbox{if $\lambda = 0$, } \\
 \sqrt{1-t} \left(\chi_{\lambda}(z) - t \chi_{\lambda-2}(z) \right) & \mbox{if $\lambda \neq 0$.} 
 \end{cases} 
\ee
The wave function for the $I_{2, 2n}$ puncture can be read off from the expression \eqref{eq:HLidxA1} to be
\be \label{eq:HLI22n}
 \psi_{\lambda}^{I_{2, 2n}} (a) = 
 \begin{cases} 
 \displaystyle
 \sqrt \frac{1+t}{1-t} & \mbox{if $\lambda=0$,}  \\
 
 \displaystyle \left( a^m + a^{-m} \right) \frac{ t^{\frac{\lambda}{2}(n+1)}}{\sqrt{1-t}}  ~~~& \mbox{otherwise.}  
 \end{cases}
\ee
Note that there are many different ways decompose the sum \eqref{eq:HLidxA1} in the form of \eqref{eq:HLtqft}. But we find our particular form given by \eqref{eq:HLI22n} is the right choice to maintain the consistent TQFT description of the index.  We find that this is indeed consistent with the accidental isomorphism of the AD theories. 

One can obtain the wave function for the $I_{2, 0}$ puncture as in the previous section by integrating the wave function for the regular puncture with the integration kernel given by vector and hypermultiplets 
\be \begin{split}
 \psi_{\lambda}^{I_{2, 0}} &= \oint [dz] I_{\textrm{vec}}(z) I_{\textrm{hyp}}(z, a) \psi_{\lambda}(z) \\
  &= \oint [dz] \PE \left[ t^{\half} (z+\frac{1}{z})(a+\frac{1}{a}) \right] P^{HL}_\lambda (z) \ , 
  \end{split}
\ee
where the wave function for the regular puncture is given as $\psi_{\lambda}(a) = \PE \left[ t \chi_{\textrm{adj}}(a) \right] P^{HL}_{\lambda}(a)$ and the vector multiplet index is given by $I_{\textrm{vec}}(z) = \PE [-t \chi_{\textrm{adj}}(z)]$. We find it agrees with the expression \eqref{eq:HLI22n} with $n=0$. 

From the TQFT structure of the index, the index for the $(A_1, D_{2n+2})$ theory can be written as
\be
 I_{(A_1, D_{2n+2})} = \sum_{\lambda} \psi^{I_{2, 2n}}_{\lambda} (x) \psi_\lambda(y)  \ .
\ee
We indeed find this expression agrees with the result \eqref{eq:A1Dn3d} from the 3d mirror. 

\paragraph{$N=2n-1$ odd}
Again, one can obtain the wave function for $I_{2, -1}$ by integrating the regular puncture wave function via vector multiplet measure. We get
\be
\begin{split}
 \psi_{\lambda}^{I_{2, -1}} &= \oint [dz] I_{\textrm{vec}}(z) \psi_{\lambda}(z) 
 = \oint [dz] P^{HL}_\lambda (z) \\
 &= 
 \begin{cases}
 	\sqrt{1-t^2} & \lambda = 0 \\
	-t \sqrt{1-t} & \lambda = 2 \\
	0 & \textrm{otherwise} \ . 
 \end{cases} 
\end{split}
\ee
From this, we obtain the HL index for the pure $SU(2)$ YM to be
\be
 I_{SYM} (t) = \sum_\lambda \psi^{I_{2, -1}}_\lambda \psi^{I_{2, -1}}_\lambda = 1-t^3 \ , 
\ee
which agrees with the direct computation. 

We can obtain the wave function for $I_{2, 1}$ from using the isomorphism $(A_1, A_3)=(A_1, D_3)$. Note that it implies
\be
 \sum_\lambda C_\lambda^{-1} \psi_\lambda^{I_{2, 4}}(x) = \sum_\lambda \psi_\lambda^{I_{2, 1}} \psi_\lambda (z) \ , 
\ee
where we identify $x = z^2$ as we learned from the Schur index. Now, we can use the orthonormality of the regular puncture wave functions. Multiply $\psi_\rho(z)$ on both sides and then integrate with the vector multiplet measure to obtain
\be
 \psi_\rho^{I_{2, 1}} = \sum_\lambda \oint [dz] I_{\textrm{vec}} (z) C_\lambda^{-1} \psi_\lambda^{I_{2, 4}}(z^2) \psi_\rho(z) 
  = \sqrt{1-t^2} \delta_{\rho, 0} \ . 
\ee
We conjecture the Hall-Littlewood wave function for $I_{2, 2n-1}$ is the same as that of $I_{2, 1}$, so that
\be
 \psi_\lambda^{I_{2, 2n-1}} = \sqrt{1-t^2} \delta_{\lambda, 0} \ , 
\ee
for $n>0$. This gives the HL index for the $(A_1, A_{2n})$ theory to be $1$, which is consistent with the absence of the Higgs branch.

%%%%%%%%%%%%%%%%%%%%%%%%%%%%%%%%%%%%%
\section{Macdonald index} \label{sec:Mac}
In this section, we discuss the Macdonald index, carrying two fugacities $(q, t)$. The Macdonald index reduces to the Schur index when $q=t$ and to the Hall-Littlewood index when $q=0$. We first construct the wave function for the puncture of type $I_{2, n}$ and then use it to write the Macdonald index for the $(A_1, A_{2n})$ theory and discuss its implications. 

\subsection{Wave function for the puncture of type $I_{2, n}$}
\paragraph{Odd $n$}
As in the previous sections, we start with the $I_{2, -1}$ puncture. It can be obtained by a `integral transformation' of a regular puncture via vector multiplet measure. The wave function for the regular puncture is given by
\be
 \psi_{\vec{\lambda}} (\vec{z}) = \PE \left[ \frac{t}{1-q} \chi_{\textrm{adj}} (\vec{z}) \right] \underline{P}_{\vec{\lambda}} (\vec{z}; q, t)  
  = \frac{1}{(t; q)^r} \prod_{\vec{\a} \in \Delta} \frac{1}{(t \vec{z}^{\vec{\a}}; q)}\underline{P}_{\vec{\lambda}} (\vec{z}; q, t)  \ , 
\ee
where $\underline{P}_{\vec \lambda} (\vec{z}; q, t) = N_{\vec{\lambda}} P_{\vec\lambda}(\vec{z}; q, t)$ is the normalized Macdonald polynomial\footnote{Our normalization is slightly different from the usual one. We include the `Cartan piece' $\frac{(q; q)^r}{(t;q)^r}$ to the measure.} such that
\be
 \langle \psi_{\vec \lambda}, \psi_{\vec \mu} \rangle = \oint [d\vec{z}]I_{\textrm{vec}}(\vec{z}) \psi_{\vec \lambda} (\vec{z}) \psi_{\vec \mu} (\vec{z}) = \delta_{\vec{\lambda} \vec{\mu}} \ .
\ee
The normalization factor for the $\G=A_1$ is given by
\be
 N_\lambda(q, t) = \left( \frac{(q; q) (t^2 q^\lambda; q)}{(t; q) (q^{\lambda+1}; q)} (1-t q^\lambda) \right)^\half = \left( \frac{(q; q)_\lambda (t^2 q^{\lambda}; q) }{(t; q)_\lambda (t q^{\lambda+1}; q)} \right)^{\half}. 
\ee
From now on we suppress the dependence on $q, t$. When $\G=A_1$, we can explicitly write down the polynomial as
\be
\underline{P}_\lambda (a) = N_\lambda \sum_{i=0}^{\lambda} \frac{(t; q)_i}{(q; q)_i} \frac{(t; q)_{\lambda-i}}{(q; q)_{\lambda-i}} a^{2i-\lambda} \ ,
\ee
where we define the $q$-Pochhammer symbol as $(x; q)_n = \prod_{i=0}^{n-1} (1-x q^i)$. The structure constant for the TQFT is given by
\be
 C_{\vec{\lambda}}^{-1} = \frac{\underline{P}_{\vec{\lambda}}(t^{\vec \rho}) }{\prod_{i=1}^r (t^{d_i}; q)} \ , 
\ee
where $\vec\rho$ is the Weyl vector of $\G =ADE$ and $d_i$ are the degrees of the Casmirs. 

Now, let us compute the wave function for the Gaiotto-Whittaker state for the $\G=A_1$ theory. We evaluate the integral to get
\be \label{eq:MacWaveftn}
\begin{split}
 \psi_\lambda^{I_{2, -1}}(q, t) &= \oint [dz] I_{\textrm{vec}} (z) \psi_{\lambda}(z; q, t)
  = \oint [dz] \PE \left(\frac{-q-t}{1-q} \chi_{\textrm{adj}}(z) \right) \PE \left[ \frac{t}{1-q} \chi_{\textrm{adj}}(z) \right] \underline{P}_{\lambda} (z)  \\
  &= N_\lambda \sum_{i=0}^{\lambda} \frac{(t; q)_i (t; q)_{\lambda-i}}{(q; q)_i (q; q)_{\lambda-i}} \oint \frac{dz}{2\pi i z} \frac{(1-z^{\pm 2})}{2} (q z^{\pm 2, 0}; q) z^{2i-\lambda} \\
  &= \begin{cases} (-1)^{\frac{\lambda}{2}} t^{\frac{\lambda}{2}} q^{\half \frac{\lambda}{2} (\frac{\lambda}{2}-1)} N_\lambda \frac{(t; q)_{\lambda/2}}{(q; q)_{\lambda/2}} &  \lambda ~ \textrm{even} ,  \\
  0 & \lambda ~ \textrm{odd} . \end{cases} 
\end{split}
\ee
We see that Schur and Hall-Littlewood limits of \eqref{eq:MacWaveftn} indeed reproduces the corresponding wave functions. 

From \eqref{eq:MacWaveftn}, we want to find a similar rescaling of the fugacities $(q, t)$ as done in the Schur and Hall-Littlewood case. 
Now, we conjecture that the wave function for $I_{2, n}$ for $n$ odd is given by
\be \label{eq:MacWaveftnN}
\psi_{\lambda}^{I_{2, n}} (q, t) = 
 \begin{cases}
% 	(-1)^{\frac{\lambda}{2}} t^{\frac{(n+2)\lambda}{2}} q^{\frac{(n+2)}{2} \frac{\lambda}{2} (\frac{\lambda}{2}-1)} \left( \frac{q}{t} \right)^{\frac{\lambda}{2} \frac{(n+1)}{2}} N_\lambda \frac{(t; q)_{\lambda/2}}{(q; q)_{\lambda/2}} &  \lambda ~ \textrm{even} , \\
(-1)^{\frac{\lambda}{2}} q^{\frac{\lambda}{2} \left( \frac{\lambda}{2} + 1 \right) \left( \frac{n}{2}+1 \right)} \left( \frac{t}{q} \right)^{\frac{\lambda}{2}\left( \frac{n+3}{2} \right)} N_\lambda \frac{(t; q)_{\lambda/2}}{(q; q)_{\lambda/2}} &  \lambda ~ \textrm{even} , \\
  0 & \lambda ~ \textrm{odd}. 
 \end{cases} 
\ee
We will provide some evidences of this proposal in the following subsection. 

\paragraph{Even $n$} 
The wave function for the $I_{2, 2n-2}$ is proposed by \cite{Buican:2015tda} as (up to normalization)
\be \label{eq:MacWaveEven}
 \psi_{\lambda}^{I_{2, 2n-2}}(x) = N_\lambda \frac{t^{n\lambda/2} q^{n\lambda^2/4}}{(t; q)} \sum_{m=0}^{\lambda} \frac{(t; q)_m(t; q)_{\lambda-m}}{(q; q)_m (q; q)_{\lambda-m}} q^{-n (\frac{\lambda}{2} - m)^2} x^{2m-\lambda} \ .
\ee
As in the previous sections, we can obtain $I_{2, 0}$ by integrating the wave function for the regular puncture with vector and hypermultiplet kernel
\be
\begin{split}
 \psi_\lambda^{I_{2, 0}} (a; q, t) = \oint [dz] I_{\textrm{vec}}(z) I_{\textrm{hyp}}(z, a) \psi_{\lambda}(z) 
 = \oint \frac{dz}{2\pi i z} \frac{(1-z^{\pm 2})}{2}  \frac{(q z^{\pm 2, 0}; q)}{(t^{\half} z^\pm a^\pm; q)} \underline{P}_{\lambda}(z; q, t) . ~~
\end{split}
\ee
We have verified that this expression agrees with \eqref{eq:MacWaveEven} for $n=1$ up to high orders in $q$. 

\subsection{Examples}
In this section, we use \eqref{eq:MacWaveftnN} to compute the Macdonald index for a number of examples and provide consistency checks. 

\subsubsection{Consistency checks}
Let us first consider a number of examples where we can cross-check the conjectured formula \eqref{eq:MacWaveftnN} against independent computations. 

\paragraph{$SU(2)$ SYM}
The pure YM theory can be realized by a pair of $I_{2, -1}$ punctures on a sphere. From the TQFT structure of the index, we write
\begin{align}
\begin{split}
 I_{SYM} &= \sum_\lambda \psi_\lambda^{I_{2, -1}} \psi_\lambda^{I_{2, -1}}
  = \sum_{n \ge 0} N_{2n}^2 \frac{(t; q)_n^2}{(q; q)_n^2} t^{2n} q^{n(n-1)} \\
  &= 1+q^2 T+q^3 (T-T^3)+q^4 (T-T^3)+q^5 (T-T^3)+q^6 (-T^3+T^2+T) \\
  &{ }~~~+q^7 (-T^4-T^3+T^2+T)+O(q^8) \ , 
\end{split} 
\end{align}
where $t = qT$. This result indeed agrees with the direct computation. 
%This example does not provide a consistency check of \eqref{eq:MacWaveftnN}, because we derived the result from the TQFT structure of the index. The following two examples provide non-trivial checks. 

\paragraph{$(A_1, A_0)$ theory}
This should describe a theory with no massless degrees of freedom. We indeed find
\be
 I_{(A_1, A_0)} = \sum_\lambda C_\lambda^{-1} \psi_\lambda^{I_{2, 1}} (q, t) = 1 \ , 
\ee
to high orders in $q$. 
This is rather a non-trivial check of the proposal \eqref{eq:MacWaveftnN}, since each term in the sum has to cancel exactly up on summing over all terms. 

\paragraph{$(A_1, A_3) = (A_1, D_3)$ theory}
The $(A_1, A_3)$ theory is isomorphic to $(A_1, D_3)$ theory. The former description can be obtained from a single $I_{2, 4}$ puncture and the latter description can be obtained from $I_{2, 1}$ and a regular puncture. We find that these two descriptions indeed give us the same index 
\be
 I_{(A_1, A_3)}(x) = \sum_\lambda C_\lambda^{-1} \psi_\lambda^{I_{2, 4}} (x) = \sum_{\lambda} \psi_{\lambda}^{I_{2, 1}} \psi_\lambda (a) = I_{(A_1, D_3)}(a) \ , 
\ee
upon identifying $x = a^2$. This result provides a consistency check between \eqref{eq:MacWaveftnN} and \eqref{eq:MacWaveEven}. 

We find that the Schur limit of this index can be written in a very simple form
\be
 I_{(A_1, D_3)}(a) = \PE \left[ \frac{q - q^3}{(1-q)(1-q^3)} \chi_{\textrm{adj}}(a) \right] \ . 
\ee
The first term inside the PE is coming from the conserved current multiplet. 

\subsubsection{Conjecture for the Macdonald indices of Argyres-Douglas theories}
\paragraph{$(A_1, A_2)$ theory}
It can be obtained from $I_{2, 3}$ punctured sphere. We conjecture its Macdonald index is given as 
\begin{align} \label{eq:a1a2mac}
\begin{split}
 I_{(A_1, A_2)} = \sum_{\lambda} C_\lambda^{-1} \psi_\lambda^{I_{2, 3}} &= 1+q^2 T+q^3 T+q^4 T+q^5 T+q^6 (T^2+T)+q^7 (T^2+T)  \\
 &{ }+ q^8 (2 T^2+T)+q^9 (2 T^2+T)+q^{10} (3 T^2+T)+O(q^{11}) \ , ~~  
\end{split}
\end{align}
where $t = q T$. This theory does not have a Higgs branch. This can be seen from triviality of the Hall-Littlewood limit of the index $q \to 0$. 

We find that this expression can be also written as
\be
 I_{(A_1, A_2)} = \PE \left[ \frac{q^2 T - q^4 T^2}{(1-q)(1-q^5 T^2)}  + O(q^{11}) \right] \ , 
\ee
where the $O(q^{11})$ terms vanish in the limit $T \to 1$. The first term inside the PE is coming from the short multiplet $\hat{\CC}_{0(0, 0)}$ (and their powers) using the notation of \cite{Dolan:2002zh}. See also appendix B of \cite{Gadde:2011uv}. This is the multiplet containing the stress-energy tensor. The Macdonald index for the short multiplet $\hat{\CC}_{R(j_1, j_2)}$ is given as
\be
 I_{\hat{\CC}_{R(j_1, j_2)}} = (-1)^{2(j_1 + j_2)} \frac{q^{2j_1 + 1} t^{R+1 + j_2 - j_1}}{1-q} = (-1)^{2(j_1 + j_2)} \frac{q^{R+2+j_1+j_2} T^{R+1+j_2-j_1}}{1-q}\ . 
\ee
We see that the stress-energy tensor multiplet contributes $\frac{q^2 T}{1-q}$ to the Macdonald index. 

Since any SCFT has a stress-energy tensor multiplet, the operator appear in the OPE of it should be also present in the theory. The OPE of the stress-energy tensor multiplet $\hat{\CC}_{0(0, 0)} \times \hat{\CC}_{0(0, 0)}$ contains \cite{Liendo:2015ofa}\footnote{The author would like to thank Wenbin Yan for discussions on this point.}
\be \label{eq:C0C0OPE}
\hat{\CC}_{0(0, 0)} \times  \hat{\CC}_{0(0, 0)} \sim \hat{\CC}_{0(\frac{\ell}{2},\frac{\ell}{2})} + \hat{\CC}_{1(\frac{\ell}{2},\frac{\ell}{2})} + \cdots \ ,
\ee
where we have only written short multiplets appear in the OPE that contributes to the Macdonald index. The $\hat{\CC}_{0(\frac{\ell}{2},\frac{\ell}{2})}$ multiplets are higher-spin conserved currents which have to be absent unless the theory is free or has a decoupled sector \cite{Maldacena:2011jn}. This multiplet contributes to the index by $\frac{q^{\ell+2} T}{1-q}$. Indeed, we see from the index \eqref{eq:a1a2mac} that there is no $\hat{\CC}_{0(\frac{\ell}{2},\frac{\ell}{2})}$ multiplet other than $\ell=0$, which contains the stress-energy tensor. 

Among the terms appear on the RHS of the OPE \eqref{eq:C0C0OPE}, $\hat{\CC}_{1(\frac{\ell}{2}, \frac{\ell}{2})}$ multiplet contributes $\frac{q^{\ell+3} T^2}{1-q}$ to the index. Since the index has coefficient $0$ for the $q^4 T^2$ term, $\hat{\CC}_{1(\half, \half)}$ cannot be present. We also see $\hat{\CC}_{1(\frac{\ell}{2}, \frac{\ell}{2})}$ with even $\ell$ is absent. Our result agrees with the analysis of \cite{Liendo:2015ofa} where they show that $\hat{\CC}_{1(\half, \half)}$ is absent for the theory with central charge $c=\frac{11}{30}$ which is the value of $(A_1, A_2)$ theory \cite{Shapere:2008zf}. See also \cite{Lemos:2015awa}.

\paragraph{$(A_1, A_4)$ theory}
It can be obtained from $I_{2, 5}$ punctured sphere. Our conjectured Macdonald index is 
\begin{align}
\begin{split}
 I_{(A_1, A_4)} = \sum_{\lambda} C_\lambda^{-1} \psi_\lambda^{I_{2, 5}} &= 1+q^2 T+q^3 T+q^4 (T^2+T)+q^5 (T^2+T)+q^6 (2 T^2+T)  \\
  &{ }+ q^7 (2 T^2+T)+q^8 (T^3+3 T^2+T)+O(q^{9}) \ . 
\end{split}
\end{align}
It also reduces to $1$ in the Hall-Littlewood limit $q \to 0$ as expected.

We find the index can be written in terms of a Plethystic exponential as
\be
 I_{(A_1, A_4)} = \PE \left[ \frac{q^2 T - q^6 T^3}{(1-q)(1-q^7T^3)} + O(q^{15}) \right] \ , 
\ee
where $O(q^{15})$ term vanishes as $T \to 1$. 

Here we see that some of the short-multiplets appear in the OPE of the 3 stress-energy tensor multiplets
\be
 \hat{\CC}_{0(0, 0)} \times \hat{\CC}_{0(0, 0)} \times \hat{\CC}_{0(0, 0)} \ , 
\ee
should be absent, because there is no term of the form $q^6 T^3$ in the index. Among the operators appear in the OPE of 3 stress-energy tensors, the multiplet contributing $\frac{q^6 T^3}{1-q}$ has to be absent. The natural candidate would be $\hat{\CC}_{2(1, 1)}$, but we cannot rule out other possibilities from the index before working out the selection rule, because any $\hat{\CC}_{R(1, j_2)}$ with $R+j_2=3$ for an integer $j_2$ will give the same index. 

\paragraph{$(A_1, A_{2n})$ theory}
We put $I_{2, 2n+1}$ puncture on a sphere. From the TQFT, we obtain the Macdonald index as
\begin{align} 
 I_{(A_1, A_{2n})}(q, t) = \sum_{\lambda} C_\lambda^{-1} \psi_\lambda^{I_{2, 2n+1}} \ . 
\end{align}
Plugging in appropriate wave functions, we get
\begin{align} \label{eq:MacIdx}
 I_{(A_1, A_{2n})}(q, t) = \frac{1}{(t^2; q)} \sum_{\lambda \in \IZ_{\ge 0}} (-1)^\lambda q^{\lambda(\lambda+1)(n+\frac{3}{2})} \left( \frac{t}{q} \right)^{\lambda(n+2)} \frac{(q^{\lambda+1})_{\lambda} (t^2 q^{2 \lambda})}{(t q^{\lambda})_{\lambda} (tq^{2\lambda+1}) } [2\lambda +1 ]_{q, t}  \ , 
\end{align}
where we used the abbreviation $(z)_n \equiv (z; q)_n$ and 
\begin{align}
 [n+1]_{q, t} = \sum_{i=0}^{n} \frac{(t; q)_i (t; q)_{n-i}}{(q; q)_i (q; q)_{n-i}} t^{\frac{n}{2} - i} \ . 
\end{align}
Note that when $t=q$, $[n]_{q, t=q} = [n]_q \equiv \frac{q^{n/2} - q^{-n/2}}{q^{1/2}-q^{-1/2}}$. In the Schur limit $t=q$, the index simplifies to the expression of \eqref{eq:A1A2nSchur}.

We find that the above index can be written as 
\be
 I_{(A_1, A_{2n})} =  \PE \left[ \frac{q^2 T - (q^2 T)^{n+1}}{(1-q)(1-q^{2n+3}T^{n+1})} + \cdots \right] \ , 
\ee
where omitted piece vanishes in the Schur limit $T \to 1$. 
There is no $(q^2 T)^{n+1}$ term in the index. Therefore the short multiplet that appear in the OPE of $(\hat{\CC}_{0(0, 0)})^{n+1}$ that contributes to the index as $\frac{(q^2 T)^{n+1}}{1-q}$ is absent. The short multiplet $\hat{\CC}_{n(\frac{n}{2}, \frac{n}{2})}$ contributes the same amount so that it might be absent.

\paragraph{$(A_1, D_5)$ theory}
It can be obtained from a sphere with a $I_{2, 3}$ puncture and a regular puncture. We get
\be \label{eq:MacA1D5}
 I_{(A_1, D_5)} (a) &=& \sum_\lambda \psi_\lambda^{I_{2, 3}} \psi_\lambda (a) \\
  &=& 1 + q T \chi_3 + q^2 \left(T (\chi_3  + \chi_1 ) + T^2 \chi_5 \right)  + q^3 \left( T(\chi_3 + \chi_1) + T^2 (\chi_5 + 2 \chi_3) + T^3 \chi_7 \right) \nn \\ 
  &{ }& + q^4 \left( T(\chi_3 + \chi_1) + T^2 (2 \chi_5 + 3 \chi_3 + 2 \chi_1) + T^3 (\chi_7 + 2 \chi_5) + T^4 \chi_9 \right) + O(q^5) \ ,  \nn
\ee
where $\chi_n$ denotes the character for the $n$-dimensional representation of $SU(2)$. 
When we take the Hall-Littlewood limit $q \to 0$ with $t$ fixed, we get
\be
 I_{(A_1, D_5)}(a) = \sum_{n \ge 0} \chi_{2n+1} (a) t^n  = \frac{1-t^2}{(1-t a^2)(1-t a^{-2})(1-t)}\ , 
\ee
which is the same as the HL index of the $(A_1, A_3)$ theory given in \eqref{eq:HLidxI22n} with $n=2$ and $w=a^2$. This is nothing but the Hilbert series of $\IC^2/\IZ_2$. 

The first term of the index \eqref{eq:MacA1D5} comes from the conserved current of the $SU(2)$ flavor symmetry. We find that the index has the form
\be
 I_{(A_1, D_5)} = \PE \left[ \frac{qT - q^5 T^3}{(1-q)(1-q^5 T^3)} \chi_{\textrm{adj}}(a) + \ldots \right] \ , 
\ee
where the omitted term vanishes in the Schur limit.\footnote{The author would like to thank Wenbin Yan for pointing out an error in the previous version.} 

\paragraph{$(A_1, D_{2n+1})$ theory}
This theory can be obtained by considering a two punctured sphere with one $I_{2, 2n-1}$ puncture and a regular puncture. Therefore the index is given by
\begin{align}
 I_{(A_1, D_{2n+1})} = \sum_{\lambda \ge 0} \psi^{I_{2, 2n-1}}_\lambda (q, t) \psi_\lambda (z; q, t) . 
\end{align}
Plugging in appropriate wave functions, we obtain
\begin{align}
 I_{(A_1, D_{2n+1})} = \frac{1}{(t z^{\pm 2, 0})_\infty} \sum_{\lambda \in \IZ_{\ge 0}} (-1)^\lambda q^{\lambda(\lambda+1)(n+\frac{1}{2})} \left(\frac{t}{q}\right)^{\lambda(n+1)} \frac{(q^{\lambda+1})_{\lambda} (t^2 q^{2 \lambda})_\infty}{(t q^{\lambda})_{\lambda} (tq^{2\lambda+1})_\infty } P_{2\lambda}(z) ,
\end{align}
where the (unnormalized) Macdonald polynomial $P_{\lambda}(z)$ is given by 
\be
P_\lambda (z) = \sum_{i=0}^{\lambda} \frac{(t; q)_i}{(q; q)_i} \frac{(t; q)_{\lambda-i}}{(q; q)_{\lambda-i}} z^{2i-\lambda} \ . 
\ee

%%%%%%%%%%%%%%%%%%%%%%%%%%%%%%%%%%%%%%%%%%%%%%%%%%
\section{$\CN=1$ class $\CS$ theories} \label{sec:N1}

For every theories in $\CN=1$ class $\CS$, the superconformal indices can be written in terms of the correlation functions of a (generalized) topological field theory on the UV curve \cite{Beem:2012yn}. See also \cite{Gadde:2013fma, Agarwal:2013uga, Agarwal:2014rua,Agarwal:2015vla}. In this section, we generalize our discussions to the $\CN=1$ case. 

\subsection{Mixed Schur index}
The $\CN=1$ index for the class $\CS$ is defined as
\be
 I (p, q, \xi; \vec{x}) = \Tr (-1)^F \fp^{j_1 + j_2 + \frac{R_0}{2}} \fq^{j_2 - j_1 + \frac{R_0}{2}} \xi^{\CF} \prod_i x_i^{F_i} \ ,  
\ee 
where $R_0$ is the UV R-charge and $\CF$ is the global $U(1)$ charge conserved for a generic class $\CS$ theory. 
One of the simplification limit of the above index is to take $\xi = \sqrt{\fq/\fp}$, called the mixed Schur limit \cite{Beem:2012yn}. It is given as
\be
 I (\fp, \fq; \vec{x}) = \Tr(-1)^F \fp^{j_1 + j_2 + J_-} \fq^{j_2 - j_1 + J_+}  \prod_i x_i^{F_i} \ , 
\ee
where we used $J_\pm = \half (R_0 \pm \CF)$. In this limit, the chiral multiplet contribution of the index can be written as
\be
 I_{\textrm{chi}} (\fp, \fq; \vec{z}) &=& \PE \left[ \frac{ \fp^{\frac{J_-}{2}} \fq^{\frac{J_+}{2}} \chi_\Lambda(\vec{z}) - \fp^{1- \frac{J_-}{2}} \fq^{1- \frac{J_+}{2}} \chi_{\bar\Lambda}(\vec{z}) }{(1-\fp)(1-\fq)} \right] \ , 
\ee
where $\chi_{\Lambda}$ is the character of the representation $\Lambda$ of the gauge group.
In the mixed Schur limit, the hypermultiplets charged with $(J_+, J_-) = (1, 0)$ gives the index a function of $\fq$ as 
\be
 I_{\rm hyp}^{(1, 0)} (\fp, \fq; \vec{z}) = \textrm{PE} \left[ \frac{\fq^\half}{1-\fq} \left( \chi_\Lambda(\vec{z})+ \chi_{\bar\Lambda}({\vec{z}}) \right) \right] \ ,
\ee
and the hypermultiplets with $(J_+, J_-) = (0, 1)$ gives 
\be
 I_{\rm hyp}^{(0, 1)} (\fp, \fq; \vec{z}) = \textrm{PE} \left[ \frac{\fp^\half}{1-\fp} ( \chi_\Lambda(\vec{z})+ \chi_{\bar\Lambda}({\vec{z}})) \right] \ .
\ee
For the chiral multiplets with $(J_+, J_-) = (0, 2)$ in the adjoint representation of $G$, we get
\be \label{eq:IdualM}
I_{\rm chi}^{(0, 2)} (\fp, \fq; \vec{z}) = \textrm{PE} \left[ \left( \frac{\fp}{1-\fp} - \frac{\fq}{1-\fq} \right) \chi_{\textrm{adj}} (\vec{z}) \right] \ ,  
\ee
where $\chi_R$ is the character of the representation $R$ of $G$. 

The index of the theory in class $\CS$ corresponding to the UV curve $\CC_{g, n}$ with normal bundle degrees $(p, q)$ is given by
\be
 I(\vec{a_i}; \fp, \fq, \xi) = \sum_{\vec \lambda} (C_{\vec \lambda}^+)^p (C_{\vec \lambda}^-)^q \prod_{i=1}^n \psi_{\vec \l}^{\s_i} (\vec{a_i})  \ , 
\ee
where $\vec\lambda$ labels the representations of $\Gamma$ and $\psi_{\vec\lambda}^\s (\vec{z})$ is the wave function associated to the puncture of color $\s=\pm$. The wave function in the mixed Schur limit becomes
\be
 \psi^+_{\vec \lambda} (\vec{z}) = \PE \left[ \frac{\fq}{1-\fq} \chi_{\textrm{adj}}(\vec{z}) \right] \chi_{\vec \lambda} (\vec{z}) \ , \quad
 \psi^-_{\vec \lambda} (\vec{z}) = \PE \left[ \frac{\fp}{1-\fp} \chi_{\textrm{adj}}(\vec{z}) \right] \chi_{\vec \lambda} (\vec{z}) \ . 
\ee
The wave functions for the irregular punctures with color $\s=\pm$ are given by simply choosing the $\CN=2$ wave function with different arguments $\psi^{I_{k, n}, +}_{\vec\lambda} = \psi^{I_{k, n}}_{\vec\lambda} (\fp)$ and  $\psi^{I_{k, n}, -}_{\vec\lambda} = \psi^{I_{k, n}}_{\vec\lambda} (\fq)$. 
In this limit, the structure constant can be also simply written as 
\be
 C_{\vec\lambda}^+ = C_{\vec\lambda}(\fq) \qquad C_{\vec\lambda}^- = C_{\vec\lambda}(\fp) \ , 
\ee
where $C_{\vec\lambda} (\fq) $ is the structure constant for $\CN=2$ theory.   

Note that one can flip the color of the wave function by attaching a chiral multiplet transforming under the adjoint of the flavor symmetry:
\be
 \psi^-_{\vec\lambda} (\vec{z}) = I_{\rm chi}^{(0, 2)}(\vec{z}) \psi^+_{\vec\lambda} (\vec{z}) \ , \quad  \psi^+_{\vec\lambda} (\vec{z}) = I_{\rm chi}^{(2, 0)}(\vec{z}) \psi^-_{\vec\lambda} (\vec{z}) \ . 
\ee
This explains why we attach flavor adjoint chiral multiplets to the oppositely colored punctures.

Equipped with the wave functions for the irregular punctures, we can easily compute the index in the mixed Schur limit. Note that even when the $\CN=2$ counterpart is non-conformal, $\CN=1$ version can actually flow to a SCFT in certain cases. For example, $SU(3)$ theory with $N_f=5$ is non-conformal for $\CN=2$, but it is in the conformal window for $\CN=1$ theory. We can indeed compute the indices for these cases using the result of section \ref{sec:Schur}. In section \ref{sec:N1su2}, we consider simplest examples. 

\subsection{$SU(2)$ SYM with $N_f=0, 1, 2, 3$} \label{sec:N1su2}
In this section, we verify that the mixed Schur index is indeed reproduced by the generalized TQFT we discussed. 

\paragraph{Pure YM}
Let us compute the index naively by using the UV matter content. Here we only have a vector multiplet. We get \cite{Spiridonov:2014cxa}
\be
 I_{YM} &=& (\fp; \fp)(\fq;\fq) \oint \frac{dz}{2\pi i z} \Delta(z) (\fq z^{\pm 2}; \fq) (\fp z^{\pm 2}; \fq) 
 = \half \sum_{m, n \in \IZ}  \fp^{\half m(m+1)} \fq^{\half n(n+1)} \oint \frac{dz}{2\pi i z} z^{2(m-n)}  \nn \\
 &=& \sum_{m \in \IZ_{\ge 0}} (\fp \fq)^{\half m(m+1)} = \sum_{\lambda} \psi^{I_{2, -1}, +}_{\lambda} \psi^{I_{2, -1}, -}_{\lambda} \ , 
\ee
where we used the Jacobi triple product identity. We indeed get the mixed Schur index from the TQFT. 

\paragraph{SQCD with $N_f=1$} 
This theory has a dynamically generated runaway superpotential, therefore we cannot define proper superconformal index. Nevertheless, we compute the index at the UV fixed point with incorrect R-charges for the chiral multiplets. Namely, we pick $R=\half$ for the chiral multiplets. This value is the correct R-charge for the mass-deformed $\CN=2$ SCFTs such as $SU(N)$ theory with $2N$ flavors. 

Now, we apply the integral formula for the index to compute
\be
 I_{N_f=1}(\fp, \fq; a) &=& (\fp; \fp)(\fq; \fq) \oint \frac{dz}{2\pi i z} \Delta(z) \frac{(\fq z^{\pm 2}; \fq) (\fp z^{\pm 2}; \fq) }{(\fq^\half z^\pm a^\pm; \fq)} \nn \\
 &=& 1+ \frac{t}{y}+ t^2 \left(-a^2-\frac{1}{a^2}+\frac{2}{y^2}\right) +t^3 \left(-\frac{a^2}{y}-\frac{1}{a^2 y}+\frac{3}{y^3}-\frac{1}{y}\right) \\
 &{ }& \qquad + t^4 \left(-\frac{2 a^2}{y^2}-\frac{2}{a^2 y^2}+\frac{5}{y^4}-\frac{1}{y^2}\right) + O(t^5) \ , \nn
\ee
where $\fp=ty, \fq=t/y$. 
This result agrees with the TQFT on a sphere with one irregular puncture $I_{2. -1}$ and one regular puncture with $(p, q)=(0, 0)$, which is given by
\be
 I_{\textrm{TQFT}} (\fp, \fq; a) = \sum_{\lambda} \psi^{I_{2, -1}, +}_{\lambda} \psi^{-}_\lambda (a) 
  = \frac{1}{(\fq a^{\pm 2, 0}; \fq)} \sum_{m \in \IZ_{\ge 0}} (-1)^m \fp^{\half m(m+1)} \chi_{R_m} (a)  \ . 
\ee
Note that we have only kept the diagonal subgroup of the full flavor symmetry $U(1)_L \times U(1)_R$.  

\paragraph{SQCD with $N_f=2$}
This theory without superpotential confines with a deformed moduli space, but we perform the computation with the same philosophy as before. Let us take the (wrong) R-charge $1/2$ to compute the index. We have two different ways to construct the theory, as we have discussed in the case of $\CN=2$ counterpart. Here depending on the choice of the colors on the punctures we actually get different indices because these choices determine the superpotential that are allowed \cite{Xie:2013rsa}. 

Let us first consider the case with two irregular punctures $I_{2, 0}$ of each color. This configuration realizes the theory with a quartic superpotential between two quarks. We write the index as
\be
 I_{N_f=1+1}(\fp, \fq; a, b) &=& (\fp; \fp)(\fq; \fq) \oint \frac{dz}{2\pi i z} \Delta(z) \frac{(\fq z^{\pm 2}; \fq) (\fp z^{\pm 2}; \fq) }{(\fq^\half z^\pm a^\pm; \fq) (\fp^\half z^\pm b^\pm; \fp) } \ , 
\ee
which agrees with the TQFT expression
\be
 I_{\textrm{TQFT}}(\fp, \fq; a, b) = \sum_{\lambda} \psi^{I_{2, 0}, +}_{\lambda} (a) \psi^{I_{2, 0}, -}_{\lambda} (b) \ .  
\ee

Now, let us consider a 3-punctured sphere realization of $N_f=2$ theory. We have two regular $+$ punctures, and one irregular $I_{2, -1}$ with $-$ color. We pick the normal bundle degrees to be $(1, 0)$. This realizes the $N_f=2$ theory without quartic superpotential, which gives the index to be
\be
 I_{N_f=2+0} (\fp, \fq; a, b) = (\fp; \fp)(\fq; \fq) \oint \frac{dz}{2\pi i z} \Delta(z) \frac{(\fq z^{\pm 2}; \fq) (\fp z^{\pm 2}; \fq) }{(\fq^\half z^\pm a^\pm; \fq) (\fq^\half z^\pm b^\pm; \fq) } \ .
\ee
It agrees with the TQFT expression
\be
 I_{\textrm{TQFT}} (\fp, \fq; x, y) = \sum_\lambda C_\lambda^{-} \psi^{I_{2, -1}, -}_\lambda \psi^{+}_\lambda (x) \psi^{+}_\lambda (y) \ ,
\ee
upon identifying $a = x y, b = x/y$.

\paragraph{SQCD with $N_f=3$}
This theory can be realized by a sphere with two regular punctures of $+$ color and one irregular puncture $I_{2, 0}$ with $-$ color, and normal bundle degrees $(p, q)=(1, 0)$. It splits $3$ flavors into $2+1$ with a quartic superpotential interaction. The index can be written as 
\be
I_{N_f=2+1} (\fp, \fq; a, b, c) = (\fp; \fp)(\fq; \fq) \oint \frac{dz}{2\pi i z} \Delta(z) \frac{(\fq z^{\pm 2}; \fq) (\fp z^{\pm 2}; \fq) }{(\fq^\half z^\pm a^\pm; \fq) (\fq^\half z^\pm b^\pm; \fq) (\fp^\half z^\pm c^\pm; \fp) } \ .
\ee
We find that it agrees with the TQFT expression
\be
 I_{\textrm{TQFT}} (\fp, \fq; x, y) = \sum_\lambda C_\lambda^{-}  \psi^{+}_\lambda (x) \psi^{+}_\lambda (y) \psi^{I_{2, 0}, -}_\lambda (c) \ ,
\ee
upon identifying $a = x y, b = x/y$.

\acknowledgments 
The author would like to thank Abhijit Gadde, Ken Intriligator, Yuji Tachikawa and Wenbin Yan for useful discussions and correspondence. The author is grateful for the hospitality of the Simons Center for Geometry and Physics during the 2015 Simons Workshop in Mathematics and Physics and also Korea Institute for Advanced Study.
This work is supported by the US Department of Energy under UCSD's contract de-sc0009919.

\bibliographystyle{jhep}
\bibliography{ADTQFT}

\providecommand{\href}[2]{#2}\begingroup\raggedright\begin{thebibliography}{10}

\bibitem{Gaiotto:2009we}
D.~Gaiotto, {\it {N=2 dualities}},  {\em JHEP} {\bf 1208} (2012) 034,
  [\href{http://xxx.lanl.gov/abs/0904.2715}{{\tt arXiv:0904.2715}}].

\bibitem{Gaiotto:2009hg}
D.~Gaiotto, G.~W. Moore, and A.~Neitzke, {\it {Wall-crossing, Hitchin Systems,
  and the WKB Approximation}},  \href{http://xxx.lanl.gov/abs/0907.3987}{{\tt
  arXiv:0907.3987}}.

\bibitem{Kinney:2005ej}
J.~Kinney, J.~M. Maldacena, S.~Minwalla, and S.~Raju, {\it {An Index for 4
  dimensional super conformal theories}},  {\em Commun. Math. Phys.} {\bf 275}
  (2007) 209--254, [\href{http://xxx.lanl.gov/abs/hep-th/0510251}{{\tt
  hep-th/0510251}}].

\bibitem{Romelsberger:2005eg}
C.~Romelsberger, {\it {Counting chiral primaries in N = 1, d=4 superconformal
  field theories}},  {\em Nucl. Phys.} {\bf B747} (2006) 329--353,
  [\href{http://xxx.lanl.gov/abs/hep-th/0510060}{{\tt hep-th/0510060}}].

\bibitem{Gadde:2009kb}
A.~Gadde, E.~Pomoni, L.~Rastelli, and S.~S. Razamat, {\it {S-duality and 2d
  Topological QFT}},  {\em JHEP} {\bf 1003} (2010) 032,
  [\href{http://xxx.lanl.gov/abs/0910.2225}{{\tt arXiv:0910.2225}}].

\bibitem{Gadde:2011ik}
A.~Gadde, L.~Rastelli, S.~S. Razamat, and W.~Yan, {\it {The 4d Superconformal
  Index from q-deformed 2d Yang-Mills}},  {\em Phys.Rev.Lett.} {\bf 106} (2011)
  241602, [\href{http://xxx.lanl.gov/abs/1104.3850}{{\tt arXiv:1104.3850}}].

\bibitem{Gadde:2011uv}
A.~Gadde, L.~Rastelli, S.~S. Razamat, and W.~Yan, {\it {Gauge Theories and
  Macdonald Polynomials}},  {\em Commun.Math.Phys.} {\bf 319} (2013) 147--193,
  [\href{http://xxx.lanl.gov/abs/1110.3740}{{\tt arXiv:1110.3740}}].

\bibitem{Gaiotto:2012xa}
D.~Gaiotto, L.~Rastelli, and S.~S. Razamat, {\it {Bootstrapping the
  superconformal index with surface defects}},  {\em JHEP} {\bf 1301} (2013)
  022, [\href{http://xxx.lanl.gov/abs/1207.3577}{{\tt arXiv:1207.3577}}].

\bibitem{Rastelli:2014jja}
L.~Rastelli and S.~S. Razamat, {\it {The superconformal index of theories of
  class $\cal S$}},  \href{http://xxx.lanl.gov/abs/1412.7131}{{\tt
  arXiv:1412.7131}}.

\bibitem{Aganagic:2004js}
M.~Aganagic, H.~Ooguri, N.~Saulina, and C.~Vafa, {\it {Black holes, q-deformed
  2d Yang-Mills, and non-perturbative topological strings}},  {\em Nucl. Phys.}
  {\bf B715} (2005) 304--348,
  [\href{http://xxx.lanl.gov/abs/hep-th/0411280}{{\tt hep-th/0411280}}].

\bibitem{Kawano:2012up}
T.~Kawano and N.~Matsumiya, {\it {5D SYM on 3D Sphere and 2D YM}},  {\em Phys.
  Lett.} {\bf B716} (2012) 450--453,
  [\href{http://xxx.lanl.gov/abs/1206.5966}{{\tt arXiv:1206.5966}}].

\bibitem{Fukuda:2012jr}
Y.~Fukuda, T.~Kawano, and N.~Matsumiya, {\it {5D SYM and 2D q-Deformed YM}},
  {\em Nucl. Phys.} {\bf B869} (2013) 493--522,
  [\href{http://xxx.lanl.gov/abs/1210.2855}{{\tt arXiv:1210.2855}}].

\bibitem{Kawano:2015ssa}
T.~Kawano and N.~Matsumiya, {\it {5D SYM on 3D Deformed Spheres}},  {\em Nucl.
  Phys.} {\bf B898} (2015) 456--562,
  [\href{http://xxx.lanl.gov/abs/1505.0656}{{\tt arXiv:1505.0656}}].

\bibitem{Benini:2011nc}
F.~Benini, T.~Nishioka, and M.~Yamazaki, {\it {4d Index to 3d Index and 2d
  TQFT}},  {\em Phys. Rev.} {\bf D86} (2012) 065015,
  [\href{http://xxx.lanl.gov/abs/1109.0283}{{\tt arXiv:1109.0283}}].

\bibitem{Alday:2013rs}
L.~F. Alday, M.~Bullimore, and M.~Fluder, {\it {On S-duality of the
  Superconformal Index on Lens Spaces and 2d TQFT}},  {\em JHEP} {\bf 05}
  (2013) 122, [\href{http://xxx.lanl.gov/abs/1301.7486}{{\tt
  arXiv:1301.7486}}].

\bibitem{Razamat:2013jxa}
S.~S. Razamat and M.~Yamazaki, {\it {S-duality and the N=2 Lens Space Index}},
  {\em JHEP} {\bf 10} (2013) 048,
  [\href{http://xxx.lanl.gov/abs/1306.1543}{{\tt arXiv:1306.1543}}].

\bibitem{Mekareeya:2012tn}
N.~Mekareeya, J.~Song, and Y.~Tachikawa, {\it {2d TQFT structure of the
  superconformal indices with outer-automorphism twists}},  {\em JHEP} {\bf
  1303} (2013) 171, [\href{http://xxx.lanl.gov/abs/1212.0545}{{\tt
  arXiv:1212.0545}}].

\bibitem{Lemos:2012ph}
M.~Lemos, W.~Peelaers, and L.~Rastelli, {\it {The superconformal index of class
  $S$ theories of type $D$}},  {\em JHEP} {\bf 1405} (2014) 120,
  [\href{http://xxx.lanl.gov/abs/1212.1271}{{\tt arXiv:1212.1271}}].

\bibitem{Chacaltana:2013oka}
O.~Chacaltana, J.~Distler, and A.~Trimm, {\it {Tinkertoys for the Twisted
  D-Series}},  {\em JHEP} {\bf 04} (2015) 173,
  [\href{http://xxx.lanl.gov/abs/1309.2299}{{\tt arXiv:1309.2299}}].

\bibitem{Agarwal:2013uga}
P.~Agarwal and J.~Song, {\it {New N=1 Dualities from M5-branes and
  Outer-automorphism Twists}},  {\em JHEP} {\bf 1403} (2014) 133,
  [\href{http://xxx.lanl.gov/abs/1311.2945}{{\tt arXiv:1311.2945}}].

\bibitem{Chacaltana:2014jba}
O.~Chacaltana, J.~Distler, and A.~Trimm, {\it {Tinkertoys for the E$_{6}$
  theory}},  {\em JHEP} {\bf 09} (2015) 007,
  [\href{http://xxx.lanl.gov/abs/1403.4604}{{\tt arXiv:1403.4604}}].

\bibitem{Chacaltana:2015bna}
O.~Chacaltana, J.~Distler, and A.~Trimm, {\it {Tinkertoys for the Twisted $E_6$
  Theory}},  \href{http://xxx.lanl.gov/abs/1501.0035}{{\tt arXiv:1501.0035}}.

\bibitem{Xie:2012hs}
D.~Xie, {\it {General Argyres-Douglas Theory}},  {\em JHEP} {\bf 1301} (2013)
  100, [\href{http://xxx.lanl.gov/abs/1204.2270}{{\tt arXiv:1204.2270}}].

\bibitem{Wang:2015mra}
Y.~Wang and D.~Xie, {\it {Classification of Argyres-Douglas theories from M5
  branes}},  \href{http://xxx.lanl.gov/abs/1509.0084}{{\tt arXiv:1509.0084}}.

\bibitem{Argyres:1995jj}
P.~C. Argyres and M.~R. Douglas, {\it {New phenomena in SU(3) supersymmetric
  gauge theory}},  {\em Nucl. Phys.} {\bf B448} (1995) 93--126,
  [\href{http://xxx.lanl.gov/abs/hep-th/9505062}{{\tt hep-th/9505062}}].

\bibitem{Argyres:1995xn}
P.~C. Argyres, M.~R. Plesser, N.~Seiberg, and E.~Witten, {\it {New N=2
  superconformal field theories in four-dimensions}},  {\em Nucl. Phys.} {\bf
  B461} (1996) 71--84, [\href{http://xxx.lanl.gov/abs/hep-th/9511154}{{\tt
  hep-th/9511154}}].

\bibitem{Buican:2015ina}
M.~Buican and T.~Nishinaka, {\it {On the Superconformal Index of
  Argyres-Douglas Theories}},  \href{http://xxx.lanl.gov/abs/1505.0588}{{\tt
  arXiv:1505.0588}}.

\bibitem{Buican:2014hfa}
M.~Buican, S.~Giacomelli, T.~Nishinaka, and C.~Papageorgakis, {\it
  {Argyres-Douglas Theories and S-Duality}},  {\em JHEP} {\bf 1502} (2015) 185,
  [\href{http://xxx.lanl.gov/abs/1411.6026}{{\tt arXiv:1411.6026}}].

\bibitem{Buican:2015hsa}
M.~Buican and T.~Nishinaka, {\it {Argyres-Douglas Theories, $S^1$ Reductions,
  and Topological Symmetries}},  \href{http://xxx.lanl.gov/abs/1505.0620}{{\tt
  arXiv:1505.0620}}.

\bibitem{Buican:2015tda}
M.~Buican and T.~Nishinaka, {\it {Argyres-Douglas Theories, the Macdonald
  Index, and an RG Inequality}},  \href{http://xxx.lanl.gov/abs/1509.0540}{{\tt
  arXiv:1509.0540}}.

\bibitem{Beem:2013sza}
C.~Beem, M.~Lemos, P.~Liendo, W.~Peelaers, L.~Rastelli, and B.~C. van Rees,
  {\it {Infinite Chiral Symmetry in Four Dimensions}},  {\em Commun. Math.
  Phys.} {\bf 336} (2015), no.~3 1359--1433,
  [\href{http://xxx.lanl.gov/abs/1312.5344}{{\tt arXiv:1312.5344}}].

\bibitem{Beem:2014rza}
C.~Beem, W.~Peelaers, L.~Rastelli, and B.~C. van Rees, {\it {Chiral algebras of
  class S}},  {\em JHEP} {\bf 1505} (2015) 020,
  [\href{http://xxx.lanl.gov/abs/1408.6522}{{\tt arXiv:1408.6522}}].

\bibitem{Lemos:2014lua}
M.~Lemos and W.~Peelaers, {\it {Chiral Algebras for Trinion Theories}},  {\em
  JHEP} {\bf 02} (2015) 113, [\href{http://xxx.lanl.gov/abs/1411.3252}{{\tt
  arXiv:1411.3252}}].

\bibitem{Cordova:2015nma}
C.~Cordova and S.-H. Shao, {\it {Schur Indices, BPS Particles, and
  Argyres-Douglas Theories}},  \href{http://xxx.lanl.gov/abs/1506.0026}{{\tt
  arXiv:1506.0026}}.

\bibitem{Cecotti:2010fi}
S.~Cecotti, A.~Neitzke, and C.~Vafa, {\it {R-Twisting and 4d/2d
  Correspondences}},  \href{http://xxx.lanl.gov/abs/1006.3435}{{\tt
  arXiv:1006.3435}}.

\bibitem{Benini:2009mz}
F.~Benini, Y.~Tachikawa, and B.~Wecht, {\it {Sicilian gauge theories and N=1
  dualities}},  {\em JHEP} {\bf 01} (2010) 088,
  [\href{http://xxx.lanl.gov/abs/0909.1327}{{\tt arXiv:0909.1327}}].

\bibitem{Bah:2012dg}
I.~Bah, C.~Beem, N.~Bobev, and B.~Wecht, {\it {Four-Dimensional SCFTs from
  M5-Branes}},  {\em JHEP} {\bf 1206} (2012) 005,
  [\href{http://xxx.lanl.gov/abs/1203.0303}{{\tt arXiv:1203.0303}}].

\bibitem{Beem:2012yn}
C.~Beem and A.~Gadde, {\it {The $N=1$ superconformal index for class $S$ fixed
  points}},  {\em JHEP} {\bf 1404} (2014) 036,
  [\href{http://xxx.lanl.gov/abs/1212.1467}{{\tt arXiv:1212.1467}}].

\bibitem{Xie:2013gma}
D.~Xie, {\it {M5 brane and four dimensional N = 1 theories I}},  {\em JHEP}
  {\bf 1404} (2014) 154, [\href{http://xxx.lanl.gov/abs/1307.5877}{{\tt
  arXiv:1307.5877}}].

\bibitem{Xie:2013jc}
D.~Xie and P.~Zhao, {\it {Central charges and RG flow of strongly-coupled N=2
  theory}},  {\em JHEP} {\bf 03} (2013) 006,
  [\href{http://xxx.lanl.gov/abs/1301.0210}{{\tt arXiv:1301.0210}}].

\bibitem{Gaiotto:2009ma}
D.~Gaiotto, {\it {Asymptotically free $\mathcal{N} = 2$ theories and irregular
  conformal blocks}},  {\em J.Phys.Conf.Ser.} {\bf 462} (2013), no.~1 012014,
  [\href{http://xxx.lanl.gov/abs/0908.0307}{{\tt arXiv:0908.0307}}].

\bibitem{Taki:2009zd}
M.~Taki, {\it {On AGT Conjecture for Pure Super Yang-Mills and W-algebra}},
  {\em JHEP} {\bf 1105} (2011) 038,
  [\href{http://xxx.lanl.gov/abs/0912.4789}{{\tt arXiv:0912.4789}}].

\bibitem{Keller:2011ek}
C.~A. Keller, N.~Mekareeya, J.~Song, and Y.~Tachikawa, {\it {The ABCDEFG of
  Instantons and W-algebras}},  {\em JHEP} {\bf 1203} (2012) 045,
  [\href{http://xxx.lanl.gov/abs/1111.5624}{{\tt arXiv:1111.5624}}].

\bibitem{Alday:2009aq}
L.~F. Alday, D.~Gaiotto, and Y.~Tachikawa, {\it {Liouville Correlation
  Functions from Four-dimensional Gauge Theories}},  {\em Lett.Math.Phys.} {\bf
  91} (2010) 167--197, [\href{http://xxx.lanl.gov/abs/0906.3219}{{\tt
  arXiv:0906.3219}}].

\bibitem{Wyllard:2009hg}
N.~Wyllard, {\it {A(N-1) conformal Toda field theory correlation functions from
  conformal N = 2 SU(N) quiver gauge theories}},  {\em JHEP} {\bf 0911} (2009)
  002, [\href{http://xxx.lanl.gov/abs/0907.2189}{{\tt arXiv:0907.2189}}].

\bibitem{Bonelli:2011aa}
G.~Bonelli, K.~Maruyoshi, and A.~Tanzini, {\it {Wild Quiver Gauge Theories}},
  {\em JHEP} {\bf 1202} (2012) 031,
  [\href{http://xxx.lanl.gov/abs/1112.1691}{{\tt arXiv:1112.1691}}].

\bibitem{Gaiotto:2012sf}
D.~Gaiotto and J.~Teschner, {\it {Irregular singularities in Liouville theory
  and Argyres-Douglas type gauge theories, I}},  {\em JHEP} {\bf 1212} (2012)
  050, [\href{http://xxx.lanl.gov/abs/1203.1052}{{\tt arXiv:1203.1052}}].

\bibitem{Kanno:2013vi}
H.~Kanno, K.~Maruyoshi, S.~Shiba, and M.~Taki, {\it {$W_3$ irregular states and
  isolated N=2 superconformal field theories}},  {\em JHEP} {\bf 1303} (2013)
  147, [\href{http://xxx.lanl.gov/abs/1301.0721}{{\tt arXiv:1301.0721}}].

\bibitem{andrews1999a}
G.~Andrews, A.~Schilling, and S.~Warnaar, {\it {An $A_2$ Bailey lemma and
  Rogers-Ramanujan-type identities}},  {\em Journal of the American
  Mathematical Society} {\bf 12} (1999), no.~3 677--702.

\bibitem{Argyres:2012fu}
P.~C. Argyres, K.~Maruyoshi, and Y.~Tachikawa, {\it {Quantum Higgs branches of
  isolated N=2 superconformal field theories}},  {\em JHEP} {\bf 10} (2012)
  054, [\href{http://xxx.lanl.gov/abs/1206.4700}{{\tt arXiv:1206.4700}}].

\bibitem{Intriligator:1996ex}
K.~A. Intriligator and N.~Seiberg, {\it {Mirror symmetry in three-dimensional
  gauge theories}},  {\em Phys.Lett.} {\bf B387} (1996) 513--519,
  [\href{http://xxx.lanl.gov/abs/hep-th/9607207}{{\tt hep-th/9607207}}].

\bibitem{2008arXiv0806.1050B}
P.~{Boalch}, {\it {Irregular connections and Kac-Moody root systems}},  {\em
  ArXiv e-prints} (June, 2008) [\href{http://xxx.lanl.gov/abs/0806.1050}{{\tt
  arXiv:0806.1050}}].

\bibitem{Cremonesi:2013lqa}
S.~Cremonesi, A.~Hanany, and A.~Zaffaroni, {\it {Monopole operators and Hilbert
  series of Coulomb branches of $3d$ $\mathcal{N} = 4$ gauge theories}},  {\em
  JHEP} {\bf 1401} (2014) 005, [\href{http://xxx.lanl.gov/abs/1309.2657}{{\tt
  arXiv:1309.2657}}].

\bibitem{Razamat:2014pta}
S.~S. Razamat and B.~Willett, {\it {Down the rabbit hole with theories of class
  S}},  \href{http://xxx.lanl.gov/abs/1403.6107}{{\tt arXiv:1403.6107}}.

\bibitem{DelZotto:2014kka}
M.~Del~Zotto and A.~Hanany, {\it {Complete Graphs, Hilbert Series, and the
  Higgs branch of the 4d N=2 $(A_n,A_m)$ SCFT's}},
  \href{http://xxx.lanl.gov/abs/1403.6523}{{\tt arXiv:1403.6523}}.

\bibitem{Gaiotto:2008ak}
D.~Gaiotto and E.~Witten, {\it {S-Duality of Boundary Conditions In N=4 Super
  Yang-Mills Theory}},  {\em Adv.Theor.Math.Phys.} {\bf 13} (2009) 721,
  [\href{http://xxx.lanl.gov/abs/0807.3720}{{\tt arXiv:0807.3720}}].

\bibitem{Maruyoshi:2013fwa}
K.~Maruyoshi, C.~Y. Park, and W.~Yan, {\it {BPS spectrum of Argyres-Douglas
  theory via spectral network}},  {\em JHEP} {\bf 12} (2013) 092,
  [\href{http://xxx.lanl.gov/abs/1309.3050}{{\tt arXiv:1309.3050}}].

\bibitem{Dolan:2002zh}
F.~Dolan and H.~Osborn, {\it {On short and semi-short representations for
  four-dimensional superconformal symmetry}},  {\em Annals Phys.} {\bf 307}
  (2003) 41--89, [\href{http://xxx.lanl.gov/abs/hep-th/0209056}{{\tt
  hep-th/0209056}}].

\bibitem{Liendo:2015ofa}
P.~Liendo, I.~Ramirez, and J.~Seo, {\it {Stress-tensor OPE in N=2
  Superconformal Theories}},  \href{http://xxx.lanl.gov/abs/1509.0003}{{\tt
  arXiv:1509.0003}}.

\bibitem{Maldacena:2011jn}
J.~Maldacena and A.~Zhiboedov, {\it {Constraining Conformal Field Theories with
  A Higher Spin Symmetry}},  {\em J. Phys.} {\bf A46} (2013) 214011,
  [\href{http://xxx.lanl.gov/abs/1112.1016}{{\tt arXiv:1112.1016}}].

\bibitem{Shapere:2008zf}
A.~D. Shapere and Y.~Tachikawa, {\it {Central charges of N=2 superconformal
  field theories in four dimensions}},  {\em JHEP} {\bf 09} (2008) 109,
  [\href{http://xxx.lanl.gov/abs/0804.1957}{{\tt arXiv:0804.1957}}].

\bibitem{Lemos:2015awa}
M.~Lemos and P.~Liendo, {\it {Bootstrapping ${\mathcal N}=2$ chiral
  correlators}},  \href{http://xxx.lanl.gov/abs/1510.0386}{{\tt
  arXiv:1510.0386}}.

\bibitem{Gadde:2013fma}
A.~Gadde, K.~Maruyoshi, Y.~Tachikawa, and W.~Yan, {\it {New N=1 Dualities}},
  {\em JHEP} {\bf 1306} (2013) 056,
  [\href{http://xxx.lanl.gov/abs/1303.0836}{{\tt arXiv:1303.0836}}].

\bibitem{Agarwal:2014rua}
P.~Agarwal, I.~Bah, K.~Maruyoshi, and J.~Song, {\it {Quiver tails and $
  \mathcal{N}=1 $ SCFTs from M5-branes}},  {\em JHEP} {\bf 1503} (2015) 049,
  [\href{http://xxx.lanl.gov/abs/1409.1908}{{\tt arXiv:1409.1908}}].

\bibitem{Agarwal:2015vla}
P.~Agarwal, K.~Intriligator, and J.~Song, {\it {Infinitely many N=1 dualities
  from $m+1-m=1$}},  \href{http://xxx.lanl.gov/abs/1505.0025}{{\tt
  arXiv:1505.0025}}.

\bibitem{Spiridonov:2014cxa}
V.~Spiridonov and G.~Vartanov, {\it {Vanishing superconformal indices and the
  chiral symmetry breaking}},  {\em JHEP} {\bf 1406} (2014) 062,
  [\href{http://xxx.lanl.gov/abs/1402.2312}{{\tt arXiv:1402.2312}}].

\bibitem{Xie:2013rsa}
D.~Xie and K.~Yonekura, {\it {Generalized Hitchin system, Spectral curve and
  $\mathcal{N} = 1$ dynamics}},  {\em JHEP} {\bf 1401} (2014) 001,
  [\href{http://xxx.lanl.gov/abs/1310.0467}{{\tt arXiv:1310.0467}}].

\end{thebibliography}\endgroup

\end{document}